# Scalable Text-Embedding-informed Cognitive Diagnosis of Large Language Models

Jia Liu    Zhiyu Xu    Yuqi Gu

Department of Statistics, Columbia University

**Abstract**

Large language models (LLMs) have achieved remarkable performance on diverse benchmarks, yet existing evaluation practices largely rely on coarse summary metrics that obscure underlying reasoning abilities. In this work, we propose novel methodologies to adapt cognitive diagnosis models (CDMs) in psychometrics to LLM evaluation, enabling fine-grained diagnosis via multidimensional discrete capability profiles and interpretable characterizations of LLM strengths and weaknesses. First, to enable CDM-based evaluation at benchmark scale (more than 1000 items), we propose a scalable method that jointly estimates LLM mastery profiles and the item-attribute **Q**-matrix, addressing key challenges posed by high-dimensional latent attributes ($K > 20$), large item pools, and the prohibitive computational cost of existing marginal maximum likelihood-based estimation. Second, we incorporate item-level textual information to construct AI-embedding-informed priors for the **Q**-matrix, stabilizing high-dimensional estimation while reducing reliance on costly human specification. We develop an efficient stochastic-approximation algorithm to jointly estimate LLM mastery profiles and the **Q**-matrix that balances data fit with text-embedding-informed priors. Simulation studies demonstrate accurate parameter recovery. An application to the MATH Level 5 benchmark illustrates the practical utility of our method for LLM evaluation and uncovers useful insights into LLMs' fine-grained capabilities.

**Keywords:** Large Language Model Evaluation; Cognitive Diagnosis Models; Embedding-Based Inference; **Q**-Matrix Learning; Parameter Estimation

# 1 Introduction

Large language models (LLMs) have demonstrated strong performance across a wide range of benchmarks, yet prevailing evaluation practices rely primarily on aggregate summary metrics that obscure the structure of underlying reasoning abilities. In response,

---

Correspondence to: Yuqi Gu. Email: `yuqi.gu@columbia.edu`. Address: Room 928, SSW Building, 1255 Amsterdam Avenue, New York, NY 10025.

recent work has moved beyond leaderboard-style comparisons toward structured, skill-level analyses of model performance (e.g., Zeng et al., 2025; Moayeri et al., 2024; Zhong et al., 2024; Zheng et al., 2026). Rather than centering evaluation solely on aggregate accuracy, a more informative perspective treats LLM assessment as the characterization of underlying capabilities that give rise to observed task performance. Accuracy summarizes outcomes, but latent abilities determine performance across heterogeneous tasks and therefore provide a more structured and multi-faceted basis for evaluation. This shift in perspective naturally motivates the use of latent variable models, as developed in psychometrics, to infer unobserved ability profiles from item-level response data.

Using with this measurement perspective, item response theory (IRT) has recently been adapted to LLM evaluation (e.g., Liu et al., 2025; Li et al., 2025; Choi et al., 2026; Castleman et al., 2025; Song et al., 2025). IRT models the interaction between model ability and item difficulty through interpretable parameters, enabling the recovery of continuous latent ability from benchmark responses. *In contrast*, cognitive diagnostic models (CDMs) characterize performance in terms of mastery or deficiency over multiple discrete skill attributes, which are linked to items through an explicit item–attribute structure known as the **Q**-matrix (Tatsuoka, 1983). This attribute-level specification provides a finer-grained and diagnostically interpretable representation of capability, allowing evaluation to move beyond global ability estimates toward structured mastery profiles.

Despite CDMs' structural advantages, adapting them to LLM evaluation introduces challenges that differ fundamentally from conventional assessment settings. LLM evaluation relies on increasingly *large benchmark suites*—often comprising thousands of heterogeneous items spanning diverse domains—creating a *benchmark scale* that departs substantially from traditional psychometric settings. For example, Big-Bench Hard (Suzgun et al., 2023) contains over 6,000 items across 23 challenging reasoning tasks, while MMLU (Hendrycks et al., 2020) includes over 15,000 multiple-choice questions spanning 57 academic subjects.

The thousands or even tens of thousands of heterogeneous items in modern bench-mark suites creates statistical and computational challenges to diagnostic evaluation of LLMs. *First*, the scale and diversity of these benchmarks enable and motivate finer-grained decomposition of abilities, resulting in high-dimensional latent attribute spaces necessary for comprehensive LLM evaluation. Since CDMs model mastery over all possible attribute patterns, the associated latent space grows exponentially with the number of attributes (i.e., $2^K$), leading to severe combinatorial complexity and rendering classical estimation procedures computationally infeasible. *Second*, **Q**-matrix specification becomes increasingly infeasible at scale. As benchmark suites expand and new tasks are continually introduced, repeated manual annotation of item–skill mappings demands substantial time and domain expertise. Existing CDM-based LLM studies have relied on expert-defined Q-matrices in domain-specific settings (Zheng et al., 2026; Kuang et al., 2025), but such manual construction is difficult to sustain in evolving large-scale benchmarks. *Third*, LLM evaluation frequently operates in regimes where the numbers of models, items, and latent attributes grow jointly. This joint growth places diagnostic modeling in high-dimensional statistical settings that depart substantially from traditional psychometric regimes. This naturally calls for reliable computational methods under such regimes, ideally with accompanying theoretical guarantees.

Motivated by these challenges, we develop new methodology for cognitive diagnosis modeling in benchmark scale settings, bringing its strengths to fine-grained LLM evaluation. Our main contributions are summarized as follows.

*Our first contribution* is a scalable and data-driven approach to **Q**-matrix learning that replaces manual expert specification with question-embedding-informed structural discovery. We leverage AI-derived embeddings of item–solution text obtained from the Qwen3-Embedding-4B model (Zhang et al., 2025) to construct a semantically organized reference **Q**-matrix, which informs a structural prior incorporated into a likelihood-based estimation framework, allowing response data to systematically validate and refine item–attribute assignments. This integration of structural priors with likelihood-based estimation yields interpretable and scalable item–skill mappings. Under this formula-tion, benchmark scale is transformed from a computational burden into a rich source of statistical signal for **Q**-matrix estimation.

As an illustration, Figure 1 visualizes the first two dimensions obtained from Uniform Manifold Approximation and Projection (UMAP; McInnes et al., 2018), a nonlinear dimensionality reduction method applied to the question–solution embeddings from the MATH Level 5 benchmark (Hendrycks et al., 2021). The benchmark provides seven official question types, which offer a coarse characterization of the items' measurement structure. Although these categories are not used to construct the embeddings, coloring the items by these labels reveals clear spatial separation in the embedding space, suggesting that the learned representations reflect meaningful structural distinctions across question types. This observation indicates that the representation is informative for uncovering attribute-level structure in the desired **Q**-matrix. Motivated by this finding, we construct an embedding-derived reference **Q**-matrix that encodes structural guidance on item groupings to guide subsequent parameter estimation. Although these official categories are coarse, later sections show that substantially finer capability structure can be recovered.

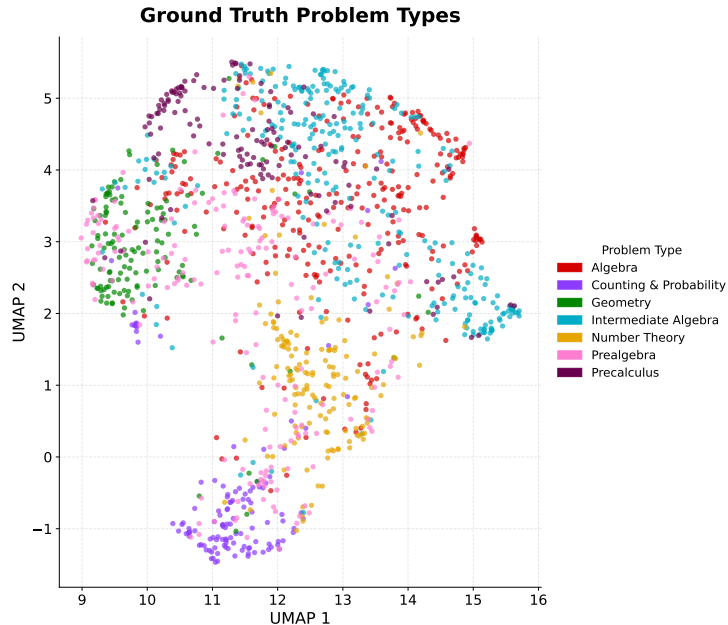

Figure 1: UMAP visualization of question–solution embeddings from the MATH-L5 benchmark, colored by official question types.

*Our second contribution* is a scalable estimation method for CDMs in regimes where both the number of items ($J$) and latent attributes ($K$) are large. When $K$ is large, the number of possible latent attribute profiles grows exponentially ($2^K$), rendering marginal maximum likelihood (MML) computationally infeasible due to high-dimensional integration. At the same time, large $J$ further amplifies the computational burden, while fully Bayesian MCMC approaches become prohibitively slow. To address this challenge, we adopt a joint estimation approach that treats latent mastery profiles as parameters and estimates them jointly with the item parameters, thereby eliminating the need to marginalize over exponentially many attribute patterns. In addition, the estimator incorporates a prior on **Q** guided by the reference **Q**-matrix, yielding a Maximum a Posteriori (MAP) estimator in which response data iteratively refine item–attribute assignments. Estimation is implemented via a stochastic approximation EM (SAEM) algorithm, enabling computational efficiency in high-dimensional settings. Under a triple-asymptotic regime in which $(N, J, K)$ diverge jointly—reflecting the practical setting of expanding benchmark suites, accumulating model releases, and increasingly fine-grained capability taxonomies—consistency of joint estimation has been established (Gu and Xu, 2023).

*Our third contribution* is a comprehensive application of our method to high-dimensional LLM data to uncover their fine-grained latent capability profiles of strengths and weaknesses. We apply the method to the MATH Level 5 benchmark (Hendrycks et al., 2021), a challenging collection of Olympiad-style mathematics problems commonly used to assess advanced mathematical reasoning in LLMs. In the processed dataset used for analysis, responses from 2,765 LLMs on 903 items are retained, providing a rich setting for studying large-scale capability structure. Using the embedding-informed procedure, we derive a well-structured and interpretable reference **Q**-matrix that organizes items into 28 semantically coherent capability groups. Comparison between the reference and posterior **Q**-matrices shows that the learned structure largely preserves this global semantic organization while introducing systematic refinements that recur in three characteristic patterns: *prior agreement, procedural augmentation, and structural reclassification.*

These recurring patterns characterize how semantic structure is validated, extended, and reorganized by item response evidence in large-scale diagnostic modeling. The resulting attribute-level mastery representations provide a structured map of skill acquisition across LLMs, supporting principled cross-base-model comparisons and revealing systematic within-family performance shifts across model versions.

The remainder of the paper is organized as follows. Section 2 presents the proposed method, including the CDM-based evaluation framework, and discuss the statistical regimes specific to LLM evaluation. Section 3 presents the embedding-informed joint estimation method, detailing **Q**-matrix learning and scalable parameter estimation. Section 4 reports results from simulation studies that assess parameter recovery and estimation performance under high-dimensional regimes. Section 5 applies the proposed methodology to real benchmark data for LLM evaluation. Finally, Section 6 concludes with discussion and future directions.

## 2 CDM-Based Evaluation of LLMs

This section develops the proposed diagnostic modeling framework for LLM evaluation. We first formalize benchmark-based evaluation within a CDM framework, characterizing model responses as manifestations of multidimensional discrete capability profiles. We then characterize the $(N, J, K)$-growth regimes that distinguish LLM evaluation from classical diagnostic settings.

### 2.1 Evaluation Framework via Cognitive Diagnosis Modeling

The evaluation of LLMs through benchmark datasets can be formulated as a latent capability measurement problem, where observed task outcomes provide indirect evidence about multidimensional discrete competencies, corresponding to latent attributes in CDMs. Each benchmark item functions as an indicator of specific capability requirements, and the resulting response pattern encodes information about an underlying mastery profile. CDMs comprise a family of models that differ in how latent attributes com-bine to generate observable outcomes, including conjunctive structures such as the DINA model (Junker and Sijtsma, 2001), disjunctive structures such as the DINO model (Templin and Henson, 2006), as well as related formulations such as the Log-Linear Cognitive Diagnosis Model (LCDM) (Henson et al., 2009) and the Generalized DINA (G-DINA) model (de la Torre, 2011).

In this work, we adopt the DINA model as the working CDM specification. Its conjunctive formulation models situations in which successful task completion depends on the joint mastery of multiple required attributes, a pattern commonly observed in complex multi-step reasoning mathematics questions. Suppose $N$ LLMs respond to a benchmark consisting of $J$ items. Let $X_{ij} \in \{0,1\}$ denote the scored response of LLM $i$ to item $j$, where $X_{ij} = 1$ indicates a correct response. Let $\boldsymbol{\alpha}_i = (\alpha_{i1}, \ldots, \alpha_{iK}) \in \{0,1\}^K$ denote the latent attribute profile across $K$ attributes. The $\mathbf{Q}$-matrix $\mathbf{Q} = (q_{jk})_{J \times K}$ encodes the item–attribute relationships, where $q_{jk} = 1$ if item $j$ requires attribute $k$, and 0 otherwise. Under the conjunctive structure, the ideal response indicator is defined as $\eta_{ij} = \prod_{k=1}^{K} \alpha_{ik}^{q_{jk}}$, which equals 1 only when individual $i$ masters all required attributes by item $j$. The DINA model specifies the probability of a correct response as

$$P_{ij} := P(X_{ij} = 1 \mid \boldsymbol{\alpha}_i, \mathbf{Q}, c_j, g_j) = c_j^{\eta_{ij}} \, g_j^{1-\eta_{ij}}, \qquad (1)$$

where $1-c_j$ and $g_j$ denote the item-specific *slipping* and *guessing* parameters, respectively. Under this parameterization, $c_j = P(X_{ij} = 1 \mid \eta_{ij} = 1)$ represents the probability of a correct response given full mastery of the required attributes, while $g_j = P(X_{ij} = 1 \mid \eta_{ij} = 0)$ represents the probability of a correct response despite lacking some of the required skills.

## 2.2 Statistical Regimes of Large-Scale LLM Evaluation

Classical cognitive diagnosis modeling is typically developed under standard asymptotic regimes in which the number of latent attributes $K$ is fixed and small, the number of items $J$ is moderate, and statistical consistency is achieved primarily through increasing sample size $N$. Under such settings, exhaustive enumeration of the $2^K$ latent mastery patterns remains computationally feasible, and **Q**-matrix specification is assumed to be externally provided or estimated within a limited-dimensional space. LLM evaluation departs fundamentally from this regime.

First, fine-grained diagnostic evaluation of LLM capabilities often requires higher attribute dimensionality (e.g., $K \geq 20$). In such settings, the space of all possible attribute profiles has cardinality $|\{0,1\}^K| = 2^K$, so marginal likelihood-based estimation entails summation over an exponentially growing configuration space. As $K$ increases, classical estimation procedures become computationally infeasible. Second, modern benchmark suites frequently contain thousands of heterogeneous items. The resulting **Q**-matrix is high-dimensional, and its specification or refinement becomes statistically unstable when both $J$ and $K$ are large. Existing exploratory and Bayesian **Q**-matrix learning procedures (e.g., Chen et al., 2018; Culpepper, 2019; Chen et al., 2015) do not scale to this regime, as they require either summation over all latent mastery configurations or joint sampling over the same exponentially large configuration space. Third, LLM evaluation naturally follows joint-growth patterns in which $N$, $J$, and $K$ may increase simultaneously as benchmarks expand and model releases accumulate. Such high-dimensional growth alters the concentration properties underlying statistical consistency.

These structural departures motivate estimation strategies that remain computationally scalable and statistically stable under high-dimensional latent discrete models with jointly increasing $(N, J, K)$. Our proposed methodology described in the subsequent sections is tailored explicitly to this regime.

## 3 Embedding-Informed Joint Estimation Framework

Building on the statistical regime characterized above, we now develop an estimation framework that incorporates structural information beyond response data alone. LLM evaluation typically provides rich meta data including item-level textual information of problem statements and reference solutions, which encodes semantic and reasoning structure not reflected in binary correctness outcomes. Empirical evidence suggests that text embedding models capture rich semantic and relational structure in natural language ([Muennighoff et al., 2023](); [Bubeck et al., 2023]()). This provides a principled basis for leveraging embeddings of benchmark items to inform **Q**-matrix construction. Advances in representation learning enable textual content to be mapped into continuous embedding spaces that preserve semantic proximity and latent conceptual organization. Such embeddings extract structural regularities relevant to item–attribute relationships. Motivated by this perspective, we propose an embedding-informed joint estimation framework for **Q**-matrix learning and parameter estimation. Specifically, embeddings are used to construct a reference **Q**-matrix encoding semantic item–attribute structure, which serves as an informed prior in the joint estimation procedure.

The remainder of this section first describes the embedding-driven construction of the reference **Q**-matrix and then develops a scalable SAEM algorithm for joint parameter estimation. Finally, we discuss the theoretical guarantees available under high-dimensional regimes that justify the use of the proposed estimator in large-scale settings.

## 3.1 Embedding-Powered Construction and Prior Specification of the Reference Q-Matrix

**Embedding-Based Structural Discovery.** To construct the reference **Q**-matrix, we require a mechanism that can uncover latent capability groupings directly from the semantic content of benchmark items. This task can be viewed as identifying coherent structure in a high-dimensional embedding space, where items with similar underlying reasoning requirements are expected to exhibit proximity in representation. To this end, we adopt and tailor the BERTopic framework ([Grootendorst, 2022]()), which combines contextual embeddings with density-based clustering to discover semantically coherent groups. Leveraging state-of-the-art paragraph embedding models, we first map each concatenated question–solution pair to a high-dimensional vector in $\mathbb{R}^D$, with $D$ typically in the order of 1000 to capture complex semantic information, forming an item embedding matrix of size $J \times D$. Following the BERTopic pipeline, non-linear dimensionality reduc-tion is applied via UMAP (McInnes et al., 2018) to obtain a reduced representation of size $J \times M$, which preserves local semantic structure while facilitating subsequent clustering.

While BERTopic employs the density-based hierarchical clustering algorithm HDBSCAN (McInnes et al., 2017), its tendency to label low-density points as outliers is incompatible with the requirement of complete item assignment in **Q**-matrix construction. To ensure full coverage of items, we apply an agglomerative hierarchical clustering procedure (Murtagh and Contreras, 2012) to the dimension-reduced embedding matrix. The resulting dendrogram can be pruned to enforce a certain minimum cluster size (e.g., 10 items), promoting interpretability and stability of the inferred attribute groupings. In addition, structural information available in the original dataset can be incorporated to guide clustering. For example, as mentioned earlier, the MATH dataset (Hendrycks et al., 2021) organizes items into seven predefined question types. This information is integrated through a distance-penalization mechanism that discourages cross-type merging during clustering. Rather than imposing a strict constraint, a large penalty factor is used to softly regularize inter-type distances, ensuring that the resulting clusters remain consistent with the predefined category structure while preserving flexibility. Consequently, the resulting clusters constitute finer-grained refinements nested within the original seven coarse categories, preserving their higher-level structure.

**Reference Q-Matrix as Structural Prior.** To interpret the latent ability represented by each cluster of items, we employ a language-model-based summarization procedure following Zeng et al. (2025). Specifically, we prompt Gemini-3-Pro to synthesize the underlying skills required by the questions and solutions within each cluster. To enhance interpretability and mitigate model-specific bias, we concurrently extract representative keywords using TF-IDF analysis (Ramos et al., 2003). When there are discrepancies between the embedding-based summary and the keyword-based evidence, we systematically examine the cluster content to refine the semantic interpretation, ensuring consistency between semantic and lexical signals.

Each resulting cluster induces a preliminary item–attribute association: items grouped within the same cluster are regarded as measuring a common latent attribute. The semantic summary and representative keywords derived from the cluster provide an interpretable label for this attribute, thereby defining an attribute name. Collectively, these cluster-level assignments define a reference $\mathbf{Q}$-matrix, denoted by $\mathbf{Q}^{(R)} = (q_{jk}^{(R)})$, where $q_{jk}^{(R)} = 1$ if item $j$ is assigned to cluster $k$, and 0 otherwise. The matrix $\mathbf{Q}^{(R)}$ serves as a structural prior in the subsequent likelihood-based estimation procedure.

## 3.2 Scalable SAEM Algorithm with a Q-Matrix Prior

We adopt a joint estimation formulation for cognitive diagnostic evaluation of LLMs, in which all unknown quantities are estimated jointly from the response data. Let $\mathbf{A} = (\boldsymbol{\alpha}_1^\top, \ldots, \boldsymbol{\alpha}_N^\top)^\top$ denote the $N \times K$ matrix collecting the latent attribute profiles of the $N$ LLMs, and let $\mathbf{c} = (c_1, \ldots, c_J)^\top$ and $\mathbf{g} = (g_1, \ldots, g_J)^\top$ denote the vectors of slipping and guessing parameters, respectively. The joint log-likelihood under the DINA model in (1) is

$$\ell(\mathbf{A}, \mathbf{Q}, \mathbf{c}, \mathbf{g} \mid \mathbf{X}) = \sum_{i=1}^{N} \sum_{j=1}^{J} \left[ X_{ij} \log \left( c_j^{\eta_{ij}} g_j^{1-\eta_{ij}} \right) + (1 - X_{ij}) \log \left( 1 - c_j^{\eta_{ij}} g_j^{1-\eta_{ij}} \right) \right],$$

where $\eta_{ij} = \prod_{k=1}^{K} \alpha_{ik}^{q_{jk}}$ denotes the ideal response indicator.

We use the embedding-derived reference matrix $\mathbf{Q}^{(R)}$ to provide structural guidance for estimating the $\mathbf{Q}$-matrix. Rather than fixing $\mathbf{Q}$ to $\mathbf{Q}^{(R)}$, we construct a prior distribution on $\mathbf{Q}$ informed by $\mathbf{Q}^{(R)}$. This formulation integrates the structural information from $\mathbf{Q}^{(R)}$ with the statistical evidence from the observed response data, allowing the estimation procedure to refine potential inaccuracies in $\mathbf{Q}^{(R)}$ arising from ambiguity in textual representations. We estimate the model parameters by maximizing the following posterior objective:

$$(\widehat{\mathbf{A}}, \widehat{\mathbf{Q}}, \widehat{\mathbf{c}}, \widehat{\mathbf{g}}) = \arg\max_{\mathbf{A}, \mathbf{Q}, \mathbf{c}, \mathbf{g}} \left\{ \ell(\mathbf{A}, \mathbf{Q}, \mathbf{c}, \mathbf{g} \mid \mathbf{X}) + \log P(\mathbf{Q} \mid \mathbf{Q}^{(R)}) \right\}. \quad (2)$$

The estimator in (2) corresponds to an MAP estimator that combines the response likelihood with the structural prior on the $\mathbf{Q}$-matrix.

Next, we develop a scalable SAEM algorithm that jointly updates the latent profiles, item parameters, and the $\mathbf{Q}$-matrix under the structural prior induced by $\mathbf{Q}^{(R)}$. At iteration $t+1$, given current estimates $\Theta^{(t)} = (\mathbf{A}^{(t)}, \mathbf{Q}^{(t)}, \mathbf{c}^{(t)}, \mathbf{g}^{(t)})$, the SAEM algorithm performs the following steps:

**S-step: Sampling of A.** For LLM $i$ and attribute $k$, we sample $\alpha_{ik}$ from its conditional posterior distribution given the current values of $\mathbf{Q}^{(t)}, \mathbf{c}^{(t)}, \mathbf{g}^{(t)}$ and the remaining attributes $\boldsymbol{\alpha}_{i,-k}$. The conditional odds can be written as

$$\Delta_{ik}^{(A)} := \log \frac{P(\alpha_{ik} = 1 \mid \boldsymbol{\alpha}_{i,-k}, \mathbf{Q}^{(t)}, \mathbf{c}^{(t)}, \mathbf{g}^{(t)}, \mathbf{X})}{P(\alpha_{ik} = 0 \mid \boldsymbol{\alpha}_{i,-k}, \mathbf{Q}^{(t)}, \mathbf{c}^{(t)}, \mathbf{g}^{(t)}, \mathbf{X})}.$$

The log-odds difference is obtained from the change in the log-likelihood when $\alpha_{ik}$ switches between 1 and 0. Let $\eta_{ij}^{(1,k)}$ and $\eta_{ij}^{(0,k)}$ denote the ideal response indicators evaluated at $\alpha_{ik} = 1$ and $\alpha_{ik} = 0$, respectively, while all other attributes remain fixed:

$$\eta_{ij}^{(1,k)} = \prod_{k' \neq k} \alpha_{ik'}^{q_{jk'}} \cdot 1^{q_{jk}}, \qquad \eta_{ij}^{(0,k)} = \prod_{k' \neq k} \alpha_{ik'}^{q_{jk'}} \cdot 0^{q_{jk}}.$$

For $m \in \{0,1\}$, define $\ell_{ij}(m) = X_{ij} \log p_j^{(m,t)} + (1 - X_{ij}) \log\left(1 - p_j^{(m,t)}\right)$, where $p_j^{(1,t)} = c_j^{(t)}$ and $p_j^{(0,t)} = g_j^{(t)}$. Then $\Delta_{ik}^{(A)}$ can be written as

$$\Delta_{ik}^{(A)} = \sum_{j: q_{jk}^{(t)} = 1} \left[ \ell_{ij}\left(\eta_{ij}^{(1,k)}\right) - \ell_{ij}\left(\eta_{ij}^{(0,k)}\right) \right].$$

From the log-odds expression above, the conditional probability can be written as $P(\alpha_{ik} = 1 \mid \boldsymbol{\alpha}_{i,-k}, \mathbf{Q}^{(t)}, \mathbf{c}^{(t)}, \mathbf{g}^{(t)}, \mathbf{X}) = \sigma(\Delta_{ik}^{(A)})$, where $\sigma(x) = 1/(1 + e^{-x})$ denotes the logistic function. Therefore, $\alpha_{ik}$ is sampled from

$$\alpha_{ik} \sim \text{Bernoulli}\left(\sigma(\Delta_{ik}^{(A)})\right). \tag{3}$$

**S-step: Sampling of Q.** For item $j$ and attribute $k$, we sample $q_{jk}$ from its conditional posterior distribution given the current values of $\mathbf{A}^{(t)}, \mathbf{c}^{(t)}, \mathbf{g}^{(t)}$ and the remaining entries $\mathbf{Q}_{-(j,k)}$. The conditional odds can be written as

$$\Delta_{jk} := \log \frac{P(q_{jk} = 1 \mid \mathbf{Q}_{-(j,k)}, \mathbf{A}^{(t)}, \mathbf{c}^{(t)}, \mathbf{g}^{(t)}, \mathbf{X})}{P(q_{jk} = 0 \mid \mathbf{Q}_{-(j,k)}, \mathbf{A}^{(t)}, \mathbf{c}^{(t)}, \mathbf{g}^{(t)}, \mathbf{X})}.$$

Under the MAP objective in (2), the log-odds decomposes into a likelihood contribution and a prior contribution: $\Delta_{jk} = \Delta_{jk}^{(\ell)} + \Delta_{jk}^{(\text{prior})}$.

The likelihood term is obtained from the change in the joint log-likelihood when $q_{jk}$ switches between 1 and 0. Let $\tilde{\eta}_{ij}^{(1,k)}$ and $\tilde{\eta}_{ij}^{(0,k)}$ denote the ideal response indicators evaluated at $q_{jk} = 1$ and $q_{jk} = 0$, respectively, while all other entries of $\mathbf{Q}$ remain fixed:

$$\tilde{\eta}_{ij}^{(1,k)} = \prod_{k' \neq k} \alpha_{ik'}^{q_{jk'}} \cdot \alpha_{ik}^1, \qquad \tilde{\eta}_{ij}^{(0,k)} = \prod_{k' \neq k} \alpha_{ik'}^{q_{jk'}} \cdot \alpha_{ik}^0.$$

Then

$$\Delta_{jk}^{(\ell)} = \sum_{i=1}^{N} \left[ \ell_{ij}(\tilde{\eta}_{ij}^{(1,k)}) - \ell_{ij}(\tilde{\eta}_{ij}^{(0,k)}) \right].$$

The prior contribution is

$$\Delta_{jk}^{(\text{prior})} = \log\left( \frac{r_{jk}}{1 - r_{jk}} \right),$$

$r_{jk} = P(q_{jk} = 1 \mid \mathbf{Q}^{(R)})$ denotes the entry-wise prior induced by the reference matrix $\mathbf{Q}^{(R)}$. Specifically, $r_{jk} = \mathbf{1}(q_{jk}^{(R)} = 1)p_{jk}^* + \mathbf{1}(q_{jk}^{(R)} = 0)(1 - p_{jk}^*)$, where $p_{jk}^*$ denotes a prior agreement probability reflecting the degree of confidence in the reference matrix. The parameter $p_{jk}^*$ may be specified entry-wise or set to a common value $p^*$.

From the log-odds expression above, the conditional probability can be written as $P(q_{jk} = 1 \mid \mathbf{Q}_{-(j,k)}, \mathbf{A}^{(t)}, \mathbf{c}^{(t)}, \mathbf{g}^{(t)}, \mathbf{X}) = \sigma(\Delta_{jk})$. Therefore, $q_{jk}$ is sampled from

$$q_{jk} \sim \text{Bernoulli}(\sigma(\Delta_{jk})).$$

**SA-step.** For each item $j$ and mastery state (ideal response state) $m \in \{0, 1\}$, define the Monte Carlo complete-data sufficient statistics

$$\widetilde{N}_j^{(m),(t+1)} = \sum_{i=1}^{N} \mathbf{1}\{\eta_{ij}^{(t+1)} = m\}, \quad \widetilde{C}_j^{(m),(t+1)} = \sum_{i=1}^{N} X_{ij} \mathbf{1}\{\eta_{ij}^{(t+1)} = m\},$$

where $\eta_{ij}^{(t+1)}$ is evaluated using the sampled latent states $(\mathbf{A}^{(t+1)}, \mathbf{Q}^{(t+1)})$ obtained in the S-step. These sufficient statistics are updated via stochastic approximation:

$$N_j^{(m),(t+1)} = (1 - \gamma_t) N_j^{(m),(t)} + \gamma_t \widetilde{N}_j^{(m),(t+1)},$$

$$C_j^{(m),(t+1)} = (1 - \gamma_t) C_j^{(m),(t)} + \gamma_t \widetilde{C}_j^{(m),(t+1)},$$

where $\{\gamma_t\}$ is a decreasing step-size sequence satisfying the standard stochastic approximation conditions $\sum_{t=1}^{\infty} \gamma_t = \infty$ and $\sum_{t=1}^{\infty} \gamma_t^2 < \infty$ (Delyon et al., 1999).

**M-step.** Given the updated sufficient statistics, the item parameters are obtained by maximizing the corresponding complete-data log-likelihood. Under the DINA model, the complete-data log-likelihood for item $j$ can be written as

$$\sum_{i=1}^{N} \Big[ \mathbf{1}\{\eta_{ij} = 1\} \big(X_{ij} \log c_j + (1 - X_{ij}) \log(1 - c_j)\big) + \mathbf{1}\{\eta_{ij} = 0\} \big(X_{ij} \log g_j + (1 - X_{ij}) \log(1 - g_j)\big) \Big]$$

$$= C_j^{(1)} \log c_j + (N_j^{(1)} - C_j^{(1)}) \log(1 - c_j) + C_j^{(0)} \log g_j + (N_j^{(0)} - C_j^{(0)}) \log(1 - g_j),$$

where $N_j^{(m)}$ and $C_j^{(m)}$ are defined in the SA-step. Maximizing this expression yields

$$c_j^{(t+1)} = \frac{C_j^{(1),(t+1)}}{N_j^{(1),(t+1)}}, \qquad g_j^{(t+1)} = \frac{C_j^{(0),(t+1)}}{N_j^{(0),(t+1)}}.$$

In particular, the algorithm scales linearly with $N$, $J$, and $K$, since the sampling updates are performed entry-wise for $\alpha_{ik}$ and $q_{jk}$ through simple Bernoulli draws, thereby avoiding enumeration of the $2^K$ latent mastery profiles. Consequently, the stochastic approximation mechanism stabilizes estimation in high-dimensional discrete spaces while preserving computational tractability.

## 3.3 Consistency

Beyond its computational scalability, the estimator also enjoys strong theoretical guarantees in the high-dimensional item-response setting. In particular, recent advances establish consistency under a triple-asymptotic regime in which the sample size $N$, the number of items $J$, and the number of latent attributes $K$ are all allowed to diverge (Gu and Xu, 2023). Within this setting, both the binary factor loadings in the **Q**-matrix and the latent attribute profiles of individuals are shown to be consistently estimable, with explicit finite-sample error bounds available for the corresponding estimators. For each item $j$ and attribute profile $\boldsymbol{\alpha} \in \{0,1\}^K$, define $p_{j,\boldsymbol{\alpha}} = P(x_{ij} = 1 \mid \boldsymbol{\alpha}_i = \boldsymbol{\alpha}, \boldsymbol{q}_j, c_j, g_j) = c_j^{\eta_j(\boldsymbol{\alpha})} g_j^{1-\eta_j(\boldsymbol{\alpha})}$, where $\eta_j(\boldsymbol{\alpha}) = \prod_{k=1}^{K} \alpha_k^{q_{jk}}$. Let $\mathbf{P} = \{p_{j,\boldsymbol{\alpha}} : j \in [J], \boldsymbol{\alpha} \in \{0,1\}^K\}$ denote the collection of item response probabilities, and let $\mathbf{P}^{\text{true}} = \{p_{j,\boldsymbol{\alpha}}^{\text{true}} : j \in [J], \boldsymbol{\alpha} \in \{0,1\}^K\}$ denote the corresponding true parameter set. The following assumptions are made on the true parameters $(\mathbf{P}^{\text{true}}, \mathbf{Q}^{\text{true}}, \mathbf{A}^{\text{true}})$ that generate the data.

**Assumption 1.** *There exists a finite number $d \geq 2$ such that*

$$\frac{1}{J^d} \leq \min_{1 \leq j \leq J,\, \boldsymbol{\alpha} \in \{0,1\}^K} p_{j,\boldsymbol{\alpha}}^{\text{true}} \leq \max_{1 \leq j \leq J,\, \boldsymbol{\alpha} \in \{0,1\}^K} p_{j,\boldsymbol{\alpha}}^{\text{true}} \leq 1 - \frac{1}{J^d}.$$

**Assumption 2.** *Suppose $c_j^{\text{true}} > g_j^{\text{true}}$ for each $j$ and that there exists a sequence $\{\beta_J\} \subset (0, \infty)$ such that $\min_{1 \leq j \leq J} \left(c_j^{\text{true}} - g_j^{\text{true}}\right)^2 \geq \beta_J$.*

**Assumption 3.** *Assume that $\sum_{k=1}^K q_{j,k}^{\text{true}} \leq K_0$ for some constant $K_0$. Also assume that there exist sequences $\{\delta_J\}, \{p_N\} \subset (0, \infty)$ and a constant $\epsilon > 0$ such that*

$$\min_{1 \leq k \leq K} \frac{1}{J} \sum_{j=1}^J \mathbb{I}(\boldsymbol{q}_j^{\text{true}} = \boldsymbol{e}_k) \geq \delta_J,$$

$$\min_{\boldsymbol{\alpha} \in \{0,1\}^K} \frac{1}{N} \sum_{i=1}^N \mathbb{I}(\boldsymbol{a}_i^{\text{true}} = \boldsymbol{\alpha}) \geq p_N \geq \frac{\epsilon}{2^K}.$$

**Theorem 1** (Consistency under the DINA model (Theorem 1 in Gu and Xu (2023))). *Consider the DINA model with $(\widehat{\mathbf{Q}}, \widehat{\mathbf{A}})$ obtained from the proposed estimation procedure. Let*

$$M = (NJ)^{-1} \sum_{i=1}^N \sum_{j=1}^J P(x_{ij} = 1 \mid \mathbf{Q}^{\text{true}}, \mathbf{A}^{\text{true}}, \mathbf{c}^{\text{true}}, \mathbf{g}^{\text{true}})$$

*denote the average positive response rate. As $N, J \to \infty$, suppose $\sqrt{J} = O(\sqrt{M} N^{1-u})$ for some $u \in (0, 1)$ and $K = o(MJ \log J)$. Under Assumptions 1–3, the following hold:*

*(a) For $\gamma_J = \frac{(\log J)^{1+\varepsilon}}{\sqrt{J}} \sqrt{M \log(2^K)}$ with some $\varepsilon > 0$, it holds that*

$$\frac{1}{NJ} \sum_{i=1}^N \sum_{j=1}^J \left(P(x_{ij} = 1 \mid \mathbf{Q}^{\text{true}}, \mathbf{A}^{\text{true}}, \mathbf{c}^{\text{true}}, \mathbf{g}^{\text{true}}) - P(x_{ij} = 1 \mid \widehat{\mathbf{Q}}, \widehat{\mathbf{A}}, \mathbf{c}^{\text{true}}, \mathbf{g}^{\text{true}})\right)^2 = o_P\left(\frac{\gamma_J}{\beta_J}\right).$$

*(b) Up to a permutation of the $K$ latent attributes,*

$$\frac{1}{J} \sum_{j=1}^J \mathbb{I}(\boldsymbol{q}_j^{\text{true}} \neq \widehat{\boldsymbol{q}}_j) = o_P\left(\frac{\gamma_J}{\beta_J p_N}\right), \quad \frac{1}{N} \sum_{i=1}^N \mathbb{I}(\widehat{\boldsymbol{a}}_i \neq \boldsymbol{a}_i^{\text{true}}) = o_P\left(\frac{\gamma_J}{\beta_J \delta_J}\right).$$

Theorem 1 establishes the recovery accuracy of the joint estimator. The stochastic fluctuation is captured by $\gamma_J$, while $\beta_J$, $p_N$, and $\delta_J$ quantify signal strength through item separation and latent profile prevalence. Consistency is guaranteed whenever the signal terms dominate the stochastic error, namely when $\gamma_J/(\beta_J p_N) \to 0$ and $\gamma_J/(\beta_J \delta_J) \to 0$, even if these signal quantities gradually decrease. Thus, the result accommodates weakening item discrimination or increasingly sparse latent profiles, provided the separation remains sufficiently strong relative to sample growth. Moreover, the theorem yields explicit finite-sample accuracy rates; in regimes where separation parameters remain bounded and $K$ is fixed, the recovery error scales on the order of $J^{-1/2}$ up to logarithmic factors.

More generally, the result indicates that consistency can be achieved in a regime where the latent dimension $K$ grows sufficiently slowly relative to the number of items $J$ and examinees $N$. This regime is consistent with modern LLM evaluation settings, where both the pool of LLMs and benchmark datasets may continue to expand over time, while the set of latent capability dimensions tends to grow more slowly or remain relatively stable, ensuring that the response information increases faster than the latent dimension. Consequently, the proposed estimation framework enjoys theoretical support for recovering latent capability structures at the benchmark scale.

It is worth noting that the optimization problem in (2) incorporates an additional prior term induced by the reference matrix $\mathbf{Q}^{(R)}$. However, when $p_{jk}^* = 0.5$, the prior becomes uninformative and the prior contribution vanishes (i.e., $\Delta_{jk}^{(\text{prior})} = 0$). In this case, the optimization reduces to the likelihood-based formulation underlying Theorem 1. Therefore, even in the absence of structural guidance from the reference $\mathbf{Q}$-matrix, the proposed framework inherits the established consistency guarantees for recovering the latent capability structure.

## 4 Simulation Study

We perform extensive simulation studies to assess the performance of the proposed SAEM algorithm in terms of recovery of the $\mathbf{Q}$-matrix and latent profiles $\mathbf{A}$, as well as estimation accuracy for the item parameters $\mathbf{c}$ and $\mathbf{g}$. We consider settings with relatively large latent dimensions, $K \in \{15, 30\}$, large item pools of size $J \in \{1000, 2000\}$, and four sample-size settings, $N \in \{500, 1000, 2000, 3000\}$, covering all three regimes in which the number of respondents is smaller than, comparable to, and larger than the number of items $J$. These configurations are designed to reflect key characteristics of LLM evaluation scenarios, where benchmarks often involve high-dimensional capability structures and large numbers of items. Two values are considered for the item parameters, with $g_j = 1 - c_j \in \{0.2, 0.3\}$ for $j = 1, \ldots, J$, as commonly adopted in simulation studies of the DINA model. For the true **Q**-matrix, we adopt a block-structured design. The $J \times K$ matrix is constructed by vertically stacking $\frac{J}{2K}$ copies of $\mathbf{I}_K$, $\frac{J}{4K}$ copies of $\mathbf{Q}^{(2)}$, and $\frac{J}{4K}$ copies of $\mathbf{Q}^{(3)}$. The block $\mathbf{Q}^{(2)}$ has entries $q_{k,k}^{(2)} = 1$ for $k = 1, \ldots, K$, $q_{k,k+1}^{(2)} = 1$ for $k = 1, \ldots, K-1$, and $q_{K,1}^{(2)} = 1$, with all other entries equal to zero. The block $\mathbf{Q}^{(3)}$ is defined by $q_{k,k}^{(3)} = 1$ for $k = 1, \ldots, K$, $q_{k,k+1}^{(3)} = 1$ for $k = 1, \ldots, K-1$, $q_{K,1}^{(3)} = 1$, $q_{k,k+2}^{(3)} = 1$ for $k = 1, \ldots, K-2$, $q_{K-1,1}^{(3)} = 1$, and $q_{K,2}^{(3)} = 1$, with all remaining entries set to zero. If $J$ is not a multiple of $K$, we first form $B = \lfloor J/K \rfloor$ full $K \times K$ blocks using the target proportions $(0.5, 0.25, 0.25)$, and then append the remaining $J \bmod K$ rows from $\mathbf{I}_K$.

To examine how the proposed algorithm leverages and refines reference information under varying levels of prior reliability, we vary the informativeness of the reference **Q**-matrix (i.e., $\mathbf{Q}^{(R)}$) along two dimensions. First, to mimic realistic settings where the reference **Q**-matrix is largely consistent with the true **Q** but differs in a small portion of entries, we construct $\mathbf{Q}^{(R)}$ by introducing random perturbations to the true **Q**-matrix. Specifically, 10% of the entries are randomly selected and replaced by independently generated Bernoulli(0.5) draws. Second, to control the strength of prior information, we specify a common prior agreement probability $p^*$ for entries of **Q**. We consider $p^* \in \{0.9, 0.6\}$, corresponding to strong and moderate prior informativeness. In total, the simulation considers $2 \times 2 \times 4 \times 2 \times 2 = 64$ distinct settings, each replicated 100 times.

Tables 1–4 summarize the recovery performance of the **Q**-matrix and latent attribute profiles **A** under different specifications of $(\mathbf{c}, \mathbf{g})$ and prior agreement probability $p^*$, measured by entry-wise and row-wise recovery rates $(\mathbf{Q}_e, \mathbf{A}_e)$ and $(\mathbf{Q}_r, \mathbf{A}_r)$. Overall, the proposed algorithm exhibits strong and stable recovery performance across all simulation settings. With $K$ and $J$ fixed, **Q**-matrix recovery improves steadily as the sample size $N$ increases, while the recovery of latent attribute profiles remains stable or improves mod-

estly. Increasing the latent dimension $K$ leads to reduced recovery accuracy for both $\mathbf{Q}$ and $\mathbf{A}$, reflecting the increased complexity of high-dimensional latent structures. Increasing the prior agreement probability from $p^* = 0.6$ to $0.9$ improves both entry-wise and row-wise recovery of the $\mathbf{Q}$-matrix, while having little effect on latent profile recovery or item parameter estimation. Finally, for fixed values of $p^*$, recovery performance improves as item parameters become less challenging, moving from $(c_j, g_j) = (0.7, 0.3)$ to $(0.8, 0.2)$.

Figures 2 and 3 display the root mean squared errors (RMSEs) of $\mathbf{g}$ and $\mathbf{c}$ as functions of $N$ under different values of $J$, $K$, and the prior agreement probability $p^*$. Rows correspond to different $(\mathbf{c}, \mathbf{g})$ settings, and columns correspond to different values of $J$. Colors indicate the latent attribute dimension $K$, while line types and symbols distinguish different values of $p^*$. Overall, the RMSEs remain small across all scenarios considered. As the sample size $N$ increases, the RMSEs of both $\mathbf{c}$ and $\mathbf{g}$ consistently decrease, indicating improved estimation accuracy with more data. In contrast, increasing the latent attribute dimension $K$, larger values of $1 - \mathbf{c}$ and $\mathbf{g}$, and decreasing $p^*$ are all associated with higher RMSEs, reflecting the increased difficulty of estimation under more challenging settings. In addition, when $K = 15$, the RMSE curves corresponding to $p^* = 0.6$ and $p^* = 0.9$ nearly overlap (see the two red curves in the figure), indicating that in lower-dimensional settings, the signal contained in the response data might be sufficient for reliable estimation, rendering performance relatively insensitive to the choice of $p^*$.

Overall, the simulation results empirically corroborate the theoretical consistency guarantees for recovering the $\mathbf{Q}$-matrix, the latent profiles, and the item parameters. In addition, the proposed algorithm enjoys superb computational scalability, with computation time scaling approximately linearly with problem size. In a challenging high-dimensional configuration with $N = 3000$, $J = 2000$, and $K = 30$, the algorithm requires on average only 6.88 minutes to complete. By contrast, such settings are typically far beyond the practical reach of existing CDM software relying on MML or Bayesian MCMC methods, which require integration over or sampling from an exponentially growing latent attribute space and thus incur prohibitive computational cost.

| $K$ | $J$ | $N$ | $\mathbf{Q}_e$ | $\mathbf{Q}_r$ | $\mathbf{A}_e$ | $\mathbf{A}_r$ |
|---|---|---|---|---|---|---|
| 15 | 1000 | 500 | 0.996 | 0.975 | 0.997 | 0.959 |
|    |      | 1000 | 1.000 | 0.999 | 0.997 | 0.961 |
|    |      | 1500 | 1.000 | 1.000 | 0.997 | 0.961 |
|    |      | 3000 | 1.000 | 1.000 | 0.997 | 0.961 |
|    | 2000 | 500 | 0.996 | 0.973 | 1.000 | 0.998 |
|    |      | 1000 | 1.000 | 0.999 | 1.000 | 0.999 |
|    |      | 1500 | 1.000 | 1.000 | 1.000 | 0.999 |
|    |      | 3000 | 1.000 | 1.000 | 1.000 | 0.999 |
| 30 | 1000 | 500 | 0.955 | 0.860 | 0.976 | 0.510 |
|    |      | 1000 | 0.995 | 0.985 | 0.982 | 0.599 |
|    |      | 1500 | 0.999 | 0.996 | 0.982 | 0.606 |
|    |      | 3000 | 1.000 | 1.000 | 0.982 | 0.612 |
|    | 2000 | 500 | 0.968 | 0.896 | 0.997 | 0.922 |
|    |      | 1000 | 0.996 | 0.989 | 0.998 | 0.938 |
|    |      | 1500 | 0.999 | 0.997 | 0.998 | 0.938 |
|    |      | 3000 | 1.000 | 1.000 | 0.998 | 0.940 |

Table 1: Recovery performance under $(c, g) = (0.7, 0.3)$ with prior agreement probability $p^* = 0.6$.

| $K$ | $J$ | $N$ | $\mathbf{Q}_e$ | $\mathbf{Q}_r$ | $\mathbf{A}_e$ | $\mathbf{A}_r$ |
|---|---|---|---|---|---|---|
| 15 | 1000 | 500 | 0.999 | 0.988 | 0.997 | 0.958 |
|    |      | 1000 | 1.000 | 1.000 | 0.997 | 0.960 |
|    |      | 1500 | 1.000 | 1.000 | 0.997 | 0.961 |
|    |      | 3000 | 1.000 | 1.000 | 0.997 | 0.961 |
|    | 2000 | 500 | 0.999 | 0.987 | 1.000 | 0.998 |
|    |      | 1000 | 1.000 | 0.999 | 1.000 | 0.999 |
|    |      | 1500 | 1.000 | 1.000 | 1.000 | 0.999 |
|    |      | 3000 | 1.000 | 1.000 | 1.000 | 0.998 |
| 30 | 1000 | 500 | 0.998 | 0.966 | 0.980 | 0.581 |
|    |      | 1000 | 1.000 | 0.998 | 0.982 | 0.602 |
|    |      | 1500 | 1.000 | 1.000 | 0.982 | 0.610 |
|    |      | 3000 | 1.000 | 1.000 | 0.982 | 0.610 |
|    | 2000 | 500 | 0.998 | 0.971 | 0.998 | 0.933 |
|    |      | 1000 | 1.000 | 0.999 | 0.998 | 0.938 |
|    |      | 1500 | 1.000 | 1.000 | 0.998 | 0.940 |
|    |      | 3000 | 1.000 | 1.000 | 0.998 | 0.940 |

Table 2: Recovery performance under $(c, g) = (0.7, 0.3)$ and prior agreement probability $p^* = 0.9$.

| $K$ | $J$ | $N$ | $\mathbf{Q}_e$ | $\mathbf{Q}_r$ | $\mathbf{A}_e$ | $\mathbf{A}_r$ |
|---|---|---|---|---|---|---|
| 15 | 1000 | 500  | 1.000 | 1.000 | 1.000 | 1.000 |
|    |      | 1000 | 1.000 | 1.000 | 1.000 | 1.000 |
|    |      | 1500 | 1.000 | 1.000 | 1.000 | 1.000 |
|    |      | 3000 | 1.000 | 1.000 | 1.000 | 1.000 |
|    | 2000 | 500  | 1.000 | 1.000 | 1.000 | 1.000 |
|    |      | 1000 | 1.000 | 1.000 | 1.000 | 1.000 |
|    |      | 1500 | 1.000 | 1.000 | 1.000 | 1.000 |
|    |      | 3000 | 1.000 | 1.000 | 1.000 | 1.000 |
| 30 | 1000 | 500  | 0.997 | 0.991 | 0.999 | 0.977 |
|    |      | 1000 | 1.000 | 0.999 | 0.999 | 0.977 |
|    |      | 1500 | 1.000 | 1.000 | 0.999 | 0.978 |
|    |      | 3000 | 1.000 | 1.000 | 0.999 | 0.978 |
|    | 2000 | 500  | 0.997 | 0.992 | 1.000 | 1.000 |
|    |      | 1000 | 1.000 | 0.999 | 1.000 | 1.000 |
|    |      | 1500 | 1.000 | 1.000 | 1.000 | 1.000 |
|    |      | 3000 | 1.000 | 1.000 | 1.000 | 1.000 |

Table 3: Recovery performance under $(c, g) = (0.8, 0.2)$ and prior agreement probability $p^* = 0.6$.

| $K$ | $J$ | $N$ | $\mathbf{Q}_e$ | $\mathbf{Q}_r$ | $\mathbf{A}_e$ | $\mathbf{A}_r$ |
|---|---|---|---|---|---|---|
| 15 | 1000 | 500  | 1.000 | 1.000 | 1.000 | 1.000 |
|    |      | 1000 | 1.000 | 1.000 | 1.000 | 1.000 |
|    |      | 1500 | 1.000 | 1.000 | 1.000 | 1.000 |
|    |      | 3000 | 1.000 | 1.000 | 1.000 | 1.000 |
|    | 2000 | 500  | 1.000 | 1.000 | 1.000 | 1.000 |
|    |      | 1000 | 1.000 | 1.000 | 1.000 | 1.000 |
|    |      | 1500 | 1.000 | 1.000 | 1.000 | 1.000 |
|    |      | 3000 | 1.000 | 1.000 | 1.000 | 1.000 |
| 30 | 1000 | 500  | 1.000 | 0.999 | 0.999 | 0.977 |
|    |      | 1000 | 1.000 | 1.000 | 0.999 | 0.978 |
|    |      | 1500 | 1.000 | 1.000 | 0.999 | 0.978 |
|    |      | 3000 | 1.000 | 1.000 | 0.999 | 0.978 |
|    | 2000 | 500  | 1.000 | 0.999 | 1.000 | 1.000 |
|    |      | 1000 | 1.000 | 1.000 | 1.000 | 1.000 |
|    |      | 1500 | 1.000 | 1.000 | 1.000 | 1.000 |
|    |      | 3000 | 1.000 | 1.000 | 1.000 | 1.000 |

Table 4: Recovery performance under $(c, g) = (0.8, 0.2)$ and prior agreement probability $p^* = 0.9$.

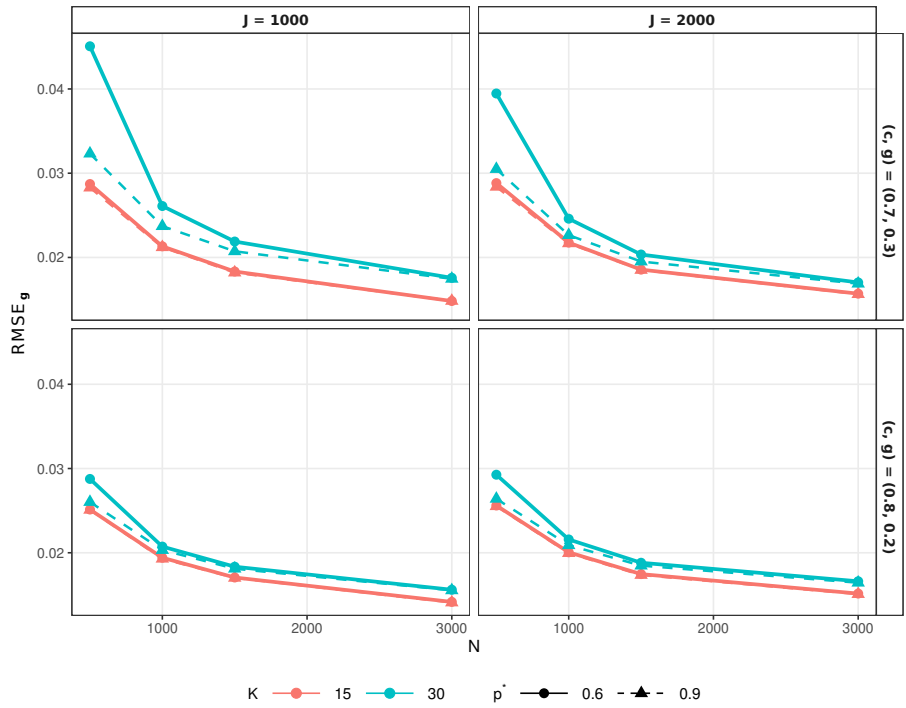

Figure 2: Root mean squared errors (RMSEs) of $g$ as functions of $N$ under different simulation settings.

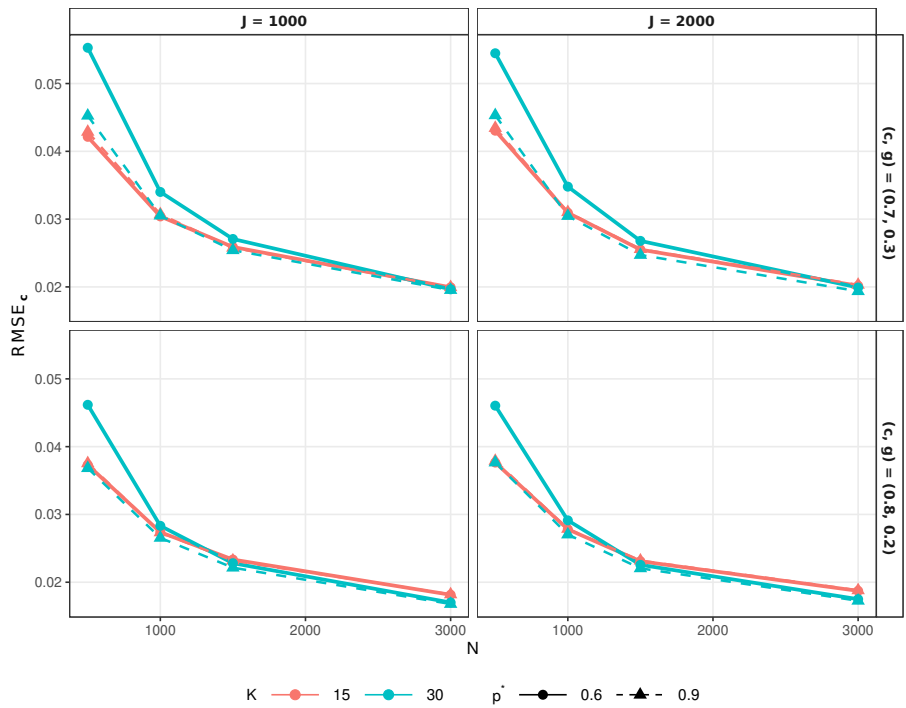

Figure 3: Root mean squared errors (RMSEs) of $c$ as functions of $N$ under different simulation settings.

# 5 Real Data Application to LLM Evaluation

To illustrate the proposed methodology in uncovering latent item–attribute structures and providing diagnostic feedback, we apply it to the MATH Level 5 dataset (Hendrycks et al., 2021), a challenging benchmark composed of Olympiad-style mathematics problems, accessed via the Hugging Face Open LLM Leaderboard repository. The original dataset contains responses from 4,491 LLMs across 1,324 items. Owing to the high difficulty of the problems, some items are rarely or never answered correctly by any LLM, while some lower-performing LLMs achieve very few correct responses across items. Such extreme response patterns provide limited diagnostic information and may adversely affect subsequent estimation. To address this issue, we preprocess the data by removing items and LLMs whose average correctness falls below 0.05. After preprocessing, the resulting dataset consists of responses from $N = 2,765$ LLMs on $J = 903$ items.

Next, we apply the proposed method to this dataset to illustrate how embedding-informed inference, together with the MAP estimator implemented via the proposed SAEM algorithm, supports principled learning of item-attribute structure and LLM latent profiles, and how the resulting structural representation enables interpretable diagnosis of LLM attribute mastery.

## 5.1 Embedding-Informed Prior and Model Fitting

Following Section 3.1, embedding-powered priors are constructed using the textual content of the questions and their solutions. We utilize Qwen3-Embedding-4B embedding model (Zhang et al., 2025), a leading embedding model on the MTEB benchmark (Muennighoff et al., 2023), to generate text embeddings. For dimension reduction, we project the embeddings into 20 dimensions using UMAP. This dimensionality is selected to be sufficiently large to preserve local structure, consistent with recommendations by McInnes et al. (2018). In the hierarchical clustering procedure, we incorporate the seven predefined question types in the MATH dataset to respect the existing metadata structure. As introduced in Section 3.1, this is implemented through the previously described distance-based regularization, which guides higher-level splits of the hierarchy to remain aligned with the broader-type categories. Consequently, the resulting clusters can be viewed as finer-grained refinements within the original seven coarse categories, preserving their overall structure while exhibiting more nuanced latent structure in the embedding space.

Within this regularized distance structure, we perform hierarchical clustering and prune the dendrogram to obtain 28 item clusters, each containing at least 10 items to ensure interpretability. These clusters induce the embedding-derived reference $\mathbf{Q}$-matrix $\mathbf{Q}^{(R)}$ used in the subsequent analysis. Detailed descriptions of these clusters are provided in Appendix Table A.1. Post-hoc verification confirms that each resulting cluster consists exclusively of questions from a single type. Figure 4 visualizes the first two UMAP dimensions of the question–solution embeddings, where the resulting clusters exhibit clear spatial separation, providing geometric evidence that the derived clusters are not only interpretable but also well differentiated.

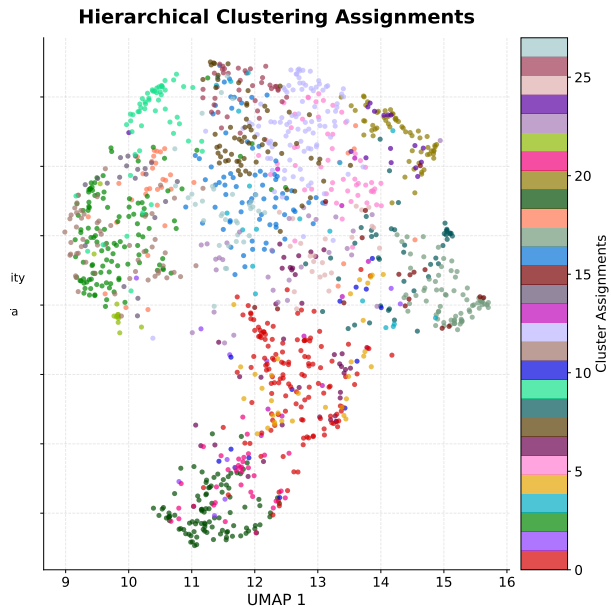

Figure 4: First two UMAP dimensions of question–solution embeddings with 28 embedding-derived clusters.

We then apply the proposed SAEM algorithm to fit the DINA model to the MATH Level 5 dataset, using $K = 28$ attributes and the embedding-derived $\mathbf{Q}^{(R)}$, with $p^* = P(q_{jk} = q_{jk}^{(R)}) = 0.9$. The estimated guessing parameters are moderate, with an average value of $\bar{g} = 0.15$, whereas the estimated slipping parameters are higher, with an average value of $\bar{c} = 0.53$. The magnitude of the estimated slipping parameters is coherent with the data context. In particular, the MATH Level 5 dataset is characterized by high overall difficulty, with an overall average correctness of 0.26 across all LLM responses. The estimated slipping parameters are strongly negatively correlated with the average item correctness, with a correlation of $-0.88$.

## 5.2  Refinement of the Item-Attribute Q-Matrix: Agreement, Augmentation, and Reclassification

For the refined **Q**-matrix, we present a subset of 45 representative items for visualization and interpretation, given the extensive scale of the full matrix. These items were strategically selected to cover all 28 latent attributes, with approximately two items associated with each attribute to maintain diagnostic representativeness. The results of the refinement process are illustrated in Figure 5. The horizontal axis denotes the 28 attribute indices, and the vertical axis lists the representative item IDs. Cells filled with light blue indicate entries of 1 in the estimated **Q**-matrix ($\widehat{q}_{ij} = 1$), while dark blue frames denote entries retained from reference **Q**-matrix ($q_{ij}^{(R)} = 1$). A fully annotated presentation of the 45 representative items—including complete problem statements, detailed solution paths, prior and refined **Q**-matrix assignments, and item-level interpretive remarks—is provided in Appendix Table A.2. Owing to the length of these materials, the appendix contains the full documentation, while Table 5 presents three selectively summarized cases that exemplify the key refinement patterns revealed by the model, which we discuss below.

Overall, the refinement results indicate that the posterior **Q**-matrix both respects the reference **Q** and introduces systematic refinements, which manifest in three recurring patterns: *prior agreement*, *procedural augmentation*, and *structural reclassification*. In high-stability cases, the refined rows of the **Q**-matrix are consistent with the reference **Q**-matrix specification. For example, in Item I1, both the prior and refined matrices identify Attribute 17 (Nonlinear Root and Complex Numbers), reflecting a prototypical application of complex arithmetic based on the cyclic properties of the imaginary unit.

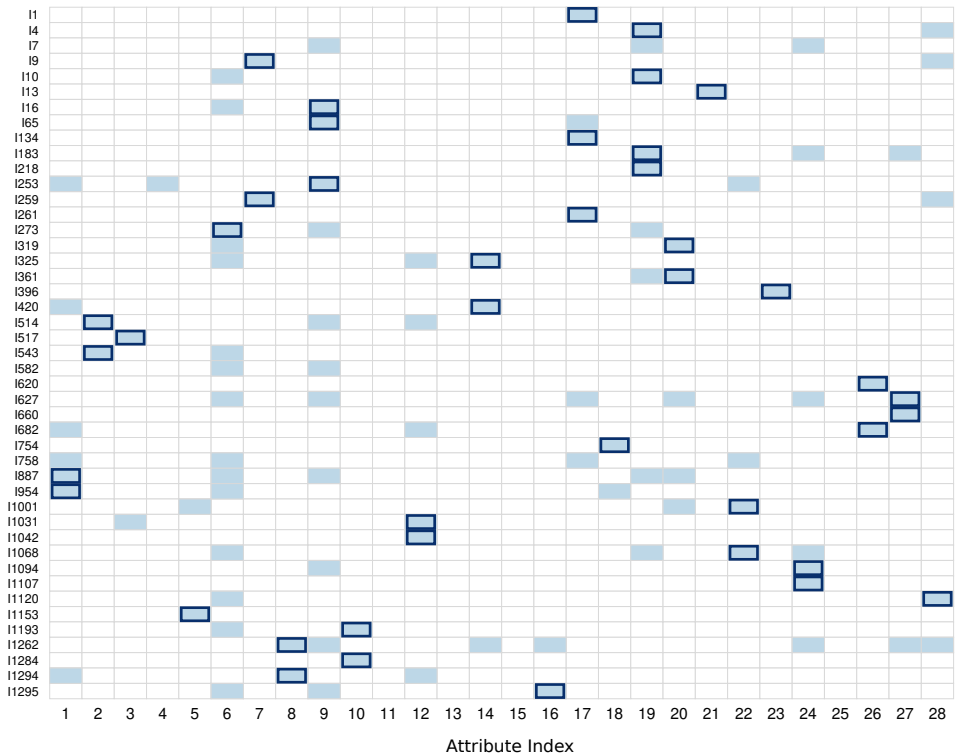

Figure 5: Comparison of the refined **Q**-matrix and reference **Q**-matrix ($N = 45$). Shaded cells indicate estimated attribute associations ($\hat{q}_{ij} = 1$), while dark blue borders denote entries recovered from the reference **Q**-matrix ($q_{ij}^{(R)} = 1$).

Beyond simple agreement, the model also performs procedural augmentation by identifying latent skills omitted in the initial reference categorization. In Item I16, while the prior assignment focused on Attribute 9 (Range of Functions), the refined **Q**-matrix additionally includes Attribute 6 (Solve Equations). This augmentation reflects the procedural necessity of solving the algebraic equation $(x^2 - 16)^2 = 0$ as part of minimizing the distance expression, thereby shifting the diagnostic emphasis toward the underlying operational requirement.

Finally, the model provides structural reclassification by refining the diagnostic emphasis suggested by the initial reference assignment. In Item I582, the prior specification assigned the item to Attribute 13 (Complex Numbers and Roots), likely influenced by the appearance of the intermediate condition $r^3 = 1$. The refined **Q**-matrix instead assigns the item to Attribute 6 (Solve Equations) together with Attribute 9 (Range of Functions). Although the derivation leads to the factorization $r^3 - 1 = (r-1)(r^2 + r + 1)$, determining the values of $a + b$ does not require explicit complex-number computation. Rather, the solution proceeds by exploiting the shared-root structure of two quadratic functions and deriving parameter constraints through algebraic manipulation. This refinement therefore reflects a shift from a surface-level categorization based on the appearance of complex-root structure toward a characterization grounded in equation solving and structural relationships among polynomial functions.

Table 5: Prototypical Cases of **Q**-matrix Refinement: Agreement, Augmentation, and Reclassification.

| ID | Question Text | Solution Path | Reference Q | Refined Q |
|---|---|---|---|---|
| I1 | **Question:** Evaluate $i^5 + i^{-25} + i^{45}$. | **Solution:** Use the cyclic property $i^n = i^{n \bmod 4}$. Hence $i^5 = i$, $i^{-25} = 1/i^{25} = -i$, and $i^{45} = i$, so the sum equals $i$. | Attr 17: Non-linear Root and Complex Numbers | Attr 17: Non-linear Root and Complex Numbers |
| I16 | **Question:** The smallest distance between the origin and a point on the graph of $y = \frac{1}{2}x^2 - 9$ can be expressed as $a$. Find $a^2$. | **Solution:** Minimize $d^2 = x^2 + y^2 = x^2 + \left(\frac{1}{2}x^2 - 9\right)^2$, equivalently minimize $\sqrt{x^2 + \frac{1}{4}x^4 - 9x^2 + 81}$. Completing the square yields $\frac{1}{2}\sqrt{(x^2 - 16)^2 + 68}$, which is minimized when $(x^2 - 16)^2 = 0$ (i.e., $x^2 = 16$). Then $a = \sqrt{17}$, so $a^2 = 17$. | Attr 9: Range of Functions | Attr 9: Range of Functions; Attr 6: Solve Equations |
| I582 | **Question:** Let $a$ and $b$ be real numbers such that $x^2 + ax + b = 0$ and $ax^2 + bx + 1 = 0$ have a root in common. Enter all possible values of $a + b$, separated by commas. | **Solution:** Let $r$ be the common root. Solving the two equations jointly implies $r^3 = 1$, so $r = 1$ or $r^2 + r + 1 = 0$. The case $r = 1$ gives $a+b = -1$; the nonreal case forces $a = b = 1$, giving $a + b = 2$. | Attr 17: Non-linear Root and Complex Numbers | Attr 6: Solve Equations; Attr 9: Range of Functions |

## 5.3 Attribute Mastery Profiles of LLMs: Strengths, Weaknesses, and Structural Gaps

We now examine the attribute-level mastery profiles of LLMs to better understand their performance across latent skills. The fitted model produces both estimated latent profile classifications and posterior mastery probabilities at the attribute level. In the following analysis, we focus on posterior mastery probabilities, as they provide a continuous mea-sure of mastery tendency and therefore facilitate more stable diagnosis and meaningful comparisons across attributes and LLMs. To begin with, Figure 6 presents the fitted density of overall averaged latent mastery alongside the density of observed accuracy, providing a global view of model performance. Although latent mastery and empirical accuracy are defined on different scales—accuracy being an observed frequency from the original dataset and latent mastery reflecting posterior expectations in the latent skill space—consistency in distributional shape and relative ordering supports the validity of the estimated mastery. Consistent with this, the Spearman rank correlation between latent mastery and observed accuracy is high ($\rho = 0.96$).

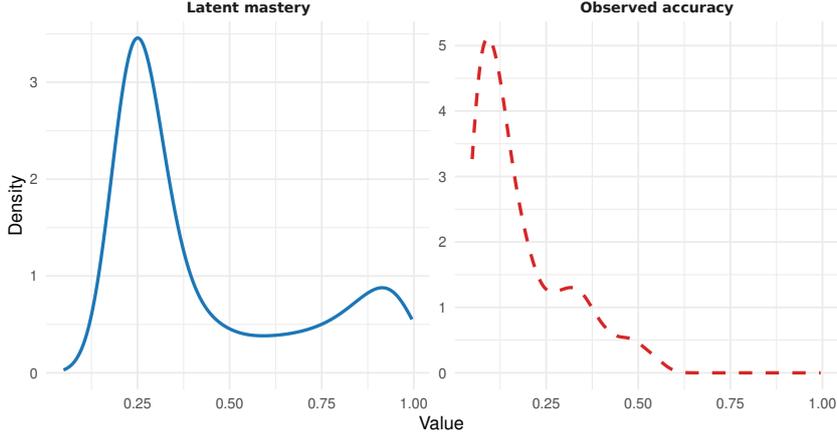

Figure 6: Distributional Comparison of Latent Mastery and Observed Accuracy. Spearman correlation: $\rho = 0.96$.

Building on the posterior mastery representation, we analyze the latent structure of LLM capabilities from complementary attribute-level and LLM-model-family perspectives, examining both skill-level constraints on performance and systematic variation associated with model version updates. At the attribute level, higher mastery is concentrated in the upper quartile of attributes, including Integer Constraints (0.50), Common Divisors or Multiples (0.50), Independence and Binomial Theorem (0.49), Triangle Inequality (0.49), Trigonometric and Binary Expansion for Optimization (0.49), Vector Operations (0.47), and Recursive Properties (0.47). In contrast, lower mastery appears in the bottom quartile, including Congruence and Modular (0.31), Circles and Angles (0.35), Nonlinear Root and Complex Numbers (0.35), Domains and Integer Counting (0.38), Word Problem with Percentage (0.38), Solve Equations (0.39), and Counting Exchangeable Objects (0.39). The remaining attributes fall between these two groups. These patterns suggest that LLMs exhibit relatively stronger performance on attributes characterized by structured algebraic manipulation and rule-based reasoning, while comparatively weaker mastery appears in attributes involving global structural, geometric, or combinatorial dependencies.

Next, we examine latent mastery profiles across LLMs. For each LLM, the posterior mastery probabilities form an attribute-level mastery vector summarizing its relative performance across the 28 attributes. These profiles provide a structured diagnostic representation at the model level. Importantly, the dataset contains metadata identifying the base model underlying each LLM. This metadata provides a principled basis for organizing LLMs into broader base model families (e.g., Gemma, LLaMA, Qwen), enabling a systematic comparison of attribute-level mastery patterns. By aggregating mastery probabilities within families, we can examine both cross-family differences and within-family variation. Figure 7 presents averaged attribute mastery profiles under two aggregation schemes. The left panel reports family-level averages across major base model families, while the right panel further disaggregates these results into sub-family profiles to reveal within-family heterogeneity, with attributes indexed along the x-axis.

Examination of family-level attribute mastery profiles reveals several clearly differentiated patterns of strengths and weaknesses across base LLM families. Qwen models exhibit consistently strong and balanced mastery across nearly all attributes, with average mastery levels typically ranging from approximately 0.55 to 0.75, indicating broad competence rather than concentration on a narrow subset of skills. DeepSeek shows relatively strong mastery in arithmetic, combinatorial, and foundational algebraic skills (e.g., common divisors or multiples at 0.58; integer partition and geometric series at 0.57; triangle inequality at 0.48), but noticeably weaker performance in linear-algebraic operations and nonlinear reasoning (e.g., vector operations at 0.20; complex numbers and roots at 0.25), resulting in a less balanced but still comparatively strong profile. In contrast,

Gemma, LLaMA, Phi, Yi, Mistral, and Mixtral exhibit broadly comparable mastery levels across many attributes, with most values falling between approximately 0.25 and 0.45. However, the degree of weakness varies by attribute and by family. For example, Mistral and Mixtral show more pronounced weakness in certain attributes, such as Domains and Integer Counting (Attribute 19) and Circles and Angles (Attribute 12), where mastery levels fall noticeably below their overall averages.

The right panel largely preserves this cross-family ranking while highlighting systematic within-family trends. In most families, mastery improves with newer model generations (e.g., from Qwen-1 to Qwen-2 and Qwen-2.5), indicating progressive gains in attribute-level competence. A notable exception is the LLaMA family, where LLaMA-2 outperforms subsequent LLaMA-3 variants on average. This pattern is also reflected in overall observed accuracy: the mean accuracy is 0.285 for LLaMA-2, compared with 0.121, 0.157, and 0.119 for LLaMA-3, LLaMA-3.1, and LLaMA-3.2, respectively. Taken together, these results demonstrate that attribute-level mastery profiles provide a structured decomposition of model performance, revealing systematic strengths, weaknesses, and structural gaps that are not visible from aggregate accuracy alone.

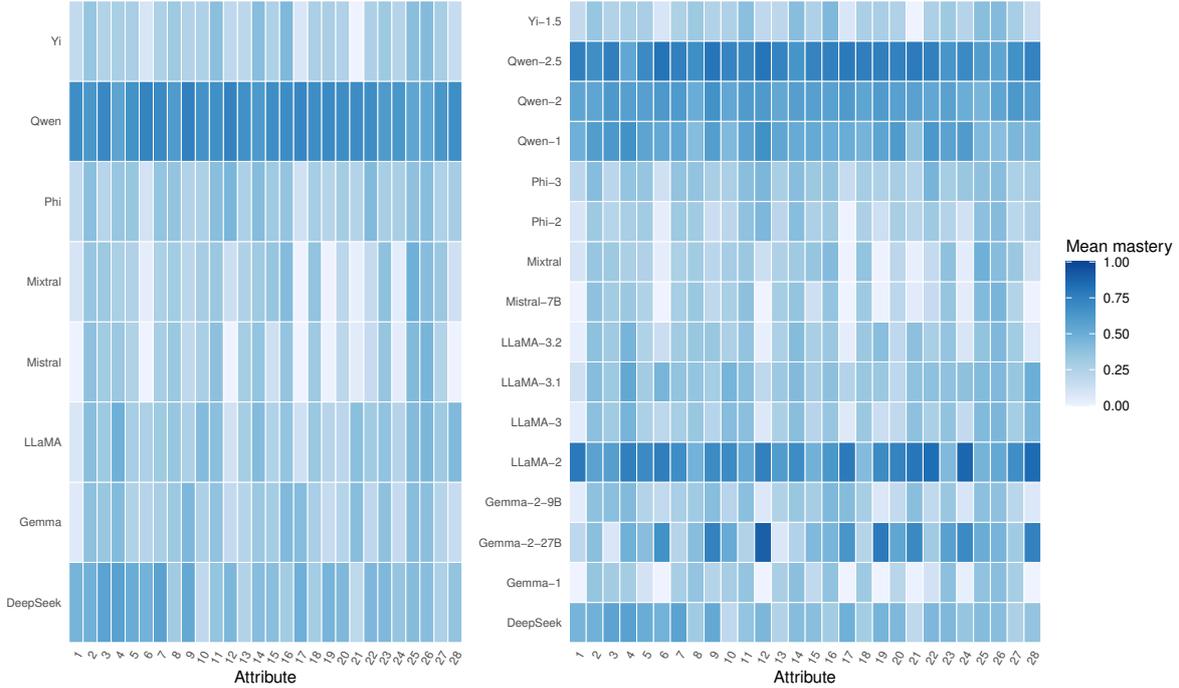

Figure 7: Averaged attribute mastery profiles under two aggregation schemes. The left panel shows base model family averages, while the right panel presents finer sub-family profiles to illustrate within-family variation.

# 6 Discussion

This work situates LLM evaluation within a fine-grained latent variable measurement framework, interpreting benchmark responses as structured evidence about multidimensional capability configurations. By statistically integrating content-informed structural information with response-driven likelihood under a joint estimation formulation, the framework reconciles representation-level information with performance-based calibration, yielding coherent and data-adaptive capability structures. Such integration is particularly important in contemporary benchmark regimes where the numbers of models, items, and latent attributes may grow jointly, requiring estimation strategies that remain both theoretically grounded and computationally tractable. The real data analysis demonstrates that this unified framework not only refines item–attribute structure in a statistically principled manner, but also supports fine-grained examination of LLM capability profiles, revealing latent skill constraints and systematic variation across model families and version updates.

An important direction for future work is the extension of the present framework to cognitive diagnostic computerized adaptive testing (CD-CAT; Chang, 2015). Given the scale of contemporary benchmarks, full evaluation across all items is often computationally burdensome and environmentally costly. Rather than relying on random item subsets, adaptive selection strategies based on information criteria—such as Kullback–Leibler divergence or Shannon entropy—could substantially reduce evaluation cost while preserving diagnostic precision. Integrating the embedding-informed estimation method with CD-CAT would allow item selection to be guided by estimated mastery uncertainty and evolving response patterns. Such extensions may follow established principles for initial item selection (Xu et al., 2016) and information-based criteria (Cheng, 2009) in cognitive diagnosis, while leveraging Monte Carlo approximations to address high-dimensional attribute spaces.

Another promising direction concerns extensions to multi-layer latent structures. For complex LLM reasoning tasks, capabilities may exhibit hierarchical organization, and modeling such structure can further enhance interpretability and representational flexibility. In this work, hierarchical information is incorporated only as known metadata (e.g., coarse question types) to guide embedding-space clustering; a natural next step is to move beyond known categories and directly model unknown higher-order latent structure within the CDM itself. This can be achieved by extending the proposed approach to identifiable deep cognitive diagnostic models (DeepCDMs; Gu, 2024; Liu and Gu, 2025), allowing fine-grained skills to be embedded within higher-order capabilities while preserving the diagnostic clarity of the current approach. Within such settings, embedding-informed priors could be constructed at multiple layers of the hierarchical **Q**-structure. Item-level representations may inform priors on fine-grained skill requirements, while higher-layer **Q**-matrices could be guided by relationships among skills derived from semantic similarity of skill descriptors. Integrating these structurally informed priors within a unified estimation framework would enable scalable modeling of hierarchical capability systems for increasingly complex LLM evaluation scenarios.

# A  Additional Results for the Real Data Analysis

This appendix provides supplementary materials for the real data analysis. Table A.1 presents the full list of the 28 derived attributes for the MATH Level 5 dataset together with their corresponding attribute IDs. Table A.2 provides a detailed breakdown of 45 representative items, including the original question–solution text, the reference and refined **Q**-matrix specifications, the refinement action relative to the reference **Q**-matrix, and the corresponding interpretation.

Table A.1: Clustered Mathematical Abilities and Attribute IDs

| Attribute ID | Ability Description |
| --- | --- |
| 1 | Congruence and Modular |
| 2 | Triangle Inequality |
| 3 | Similarity and Angle Bisector Theorem |
| 4 | Common Divisors or Multiples |
| 5 | Divisability |
| 6 | Solve Equations |
| 7 | Integer Partition and Geometric Series |
| 8 | Matrix Determinant and Trigonometry |
| 9 | Range of Functions |
| 10 | Vector Operations |
| 11 | Integer Constraints |
| 12 | Circles and Angles |
| 13 | Complex Numbers and Roots |
| 14 | Independence and Binomial Theorem |
| 15 | Conic Sections |
| 16 | Trigonometric and Binary Expansion for Optimization |
| 17 | Nonlinear Root and Complex Numbers |
| 18 | Inequality for Optimization |
| 19 | Domains and Integer Counting |
| 20 | Counting Exchangeable Objects |
| 21 | Parabola and Discriminant |
| 22 | Counting Frequency with Constraints / Addition or Multiplication Rule of Counting |
| 23 | Continuous Probability and Geometric Interpretation |
| 24 | Word Problem with Percentage |
| 25 | Tangency and Domain of Composite Functions |
| 26 | Recursive Properties |
| 27 | Polynomial Functions and Root |
| 28 | Game Theory with Integers |

Table A.2: Comprehensive breakdown of representative items ($N = 45$) with verbatim original text and refined **Q**-matrix transitions.

| Id | Item Full Text (Original Question & Solution) | Reference Q | Refined Q | Action | Mathematical Interpretation |
|---|---|---|---|---|---|
| I1 | **Question:** Evaluate $i^5 + i^{-25} + i^{45}$. **Solution:** We have $i^5 = i^4 \cdot i = 1 \cdot (i) = i$. We also have $i^{-25} = 1/i^{25} = 1/(i^{24} \cdot i) = 1/[1 \cdot (i)] = 1/i = \frac{1}{i} \cdot \frac{i}{i} = i/(-1) = -i$ and $i^{45} = (i^{44}) \cdot i = 1 \cdot i = i$, and . So, adding these three results gives $i^5 + i^{-25} + i^{45} = i + -i + i = \boxed{i}$. | 17: Non-linear Root and Complex Numbers | 17: Non-linear Root and Complex Numbers | None | Validates model stability in identifying the cyclic nature of imaginary unit powers: $i^n = i^{n \pmod 4}$. |
| I4 | **Question:** What is the smallest real number $x$ in the domain of the function $$g(x) = \sqrt{(x-3)^2 - (x-8)^2} \ ?$$ **Solution:** A real number $x$ is in the domain of $g$ if and only if $$(x-3)^2 - (x-8)^2 \geq 0.$$ Expanding this out and simplifying, we get $$10x - 55 \geq 0;$$ the smallest solution is $x = \frac{55}{10} = \boxed{\frac{11}{2}}$. Alternatively, once we have the quadratic equation $$(x-3)^2 - (x-8)^2 \geq 0,$$ instead of expanding it out, we can observe that $(x-3)^2$ is the square of the distance from $x$ to 3 on the number line, while $(x-8)^2$ is the square of the distance from $x$ to 8. Thus, $(x-3)^2 - (x-8)^2 \geq 0$ is true if $x$ is closer to 8 than to 3, which is true if and only if $x \geq \frac{8+3}{2} = \boxed{\frac{11}{2}}$. | 19: Domains and Integer Counting | 19: Domains and Integer Counting 28: Game Theory with Integers | Added: 28 | Captures the "Alternative Solution's" reliance on strategic distance logic $\|x-3\| \leq \|x-8\|$ rather than pure algebraic expansion. |

Continued on next page.

| Id | Item Full Text (Original Question & Solution) | Prior Q | Refined Q | Action | Mathematical Interpretation |
|---|---|---|---|---|---|
| I7 | **Question:** Find all positive integer values of $c$ such that the equation $x^2 - 7x + c = 0$ only has roots that are real and rational. Express them in decreasing order, separated by commas.<br>**Solution:** For the roots to be real and rational, the discriminant must be a perfect square. The discriminant is $D = (-7)^2 - 4(1)(c) = 49 - 4c$. Since $c$ is a positive integer, $49 - 4c < 49$. Also, $4c$ is a multiple of 4, so $49 - 4c$ is 1 more than a multiple of 4. The perfect squares less than 49 are 36, 25, 16, 9, 4, 1, 0. Of these, only 25, 9, and 1 are 1 more than a multiple of 4.<br>If $49 - 4c = 25$, then $4c = 24$, so $c = 6$.<br>If $49 - 4c = 9$, then $4c = 40$, so $c = 10$.<br>If $49 - 4c = 1$, then $4c = 48$, so $c = 12$.<br>So the possible values of $c$ are $\boxed{12, 10, 6}$. | 21: Parabola and Discriminant | 9: Range of Functions<br>19: Domains and Integer Counting<br>21: Parabola and Discriminant | Added: 9, 19 | Supplements the discriminant logic $\Delta = k^2$ (Attr 21) with integer constraints $c \in \mathbb{Z}^+$ (Attr 19) and range bounds (Attr 9). |
| I9 | **Question:** John computes the sum of the elements of each of the 15 two-element subsets of $\{1, 2, 3, 4, 5, 6\}$. What is the sum of these 15 sums?<br>**Solution:** Among the two-element subsets of $\{1, 2, 3, 4, 5, 6\}$, each element in $\{1, 2, 3, 4, 5, 6\}$ appears 5 times, one time in the same subset with each other element. Thus, the desired sum is $5(1+2+3+4+5+6) = 5\left(\frac{6 \cdot 7}{2}\right) = \boxed{105}$. | 7: Integer Partition and Geometric Series | 7: Integer Partition and Geometric Series<br>28: Game Theory with Integers | Added: 28 | Maps the "symmetry strategy" (frequency count $n-1$) to Integrated Logic (Attr 28), treating the avoidance of brute-force calculation as a strategic reasoning skill. |

| Id | Item Full Text (Original Question & Solution) | Prior Q | Refined Q | Action | Mathematical Interpretation |
|---|---|---|---|---|---|
| I10 | **Question:** The points $(x, y)$ represented in this table lie on a straight line. The point $(13, q)$ lies on the same line. What is the value of $p+q$? Express your answer as a decimal to the nearest tenth. $$\begin{array}{c|c} x & y \\ \hline 2 & -5 \\ p & -14 \\ p+2 & -17 \end{array}$$ **Solution:** If we have two points $(x_1, y_1)$ and $(x_2, y_2)$ on a line, we can find the slope of the line by using the formula $\dfrac{y_1 - y_2}{x_1 - x_2}$. So, for the line we are given, the slope is $\dfrac{(-5) - (-14)}{2 - p} = \dfrac{9}{2 - p}$, and the slope is also $\dfrac{(-14) - (-17)}{p - (p+2)} = \dfrac{3}{-2}$. Setting these values equal, we get $(2-p)(3) = (-2)(9)$ $6 - 3p = -18$ $p = 8.$ Now we need to find $q$. Using the same strategy as above, $\dfrac{q - (-5)}{13 - 2} = \dfrac{3}{-2}$ we find that $(11)(3) = (-2)(q+5)$ $33 = -2q - 10$ $q = -21.5.$ Therefore, $p+q = 8 + (-21.5) = \boxed{-13.5}$. | 19: Domains and Integer Counting | 6: Solve Equations 19: Domains and Integer Counting | Added: 6 | Correctly identifies the extensive algebraic steps required to solve the system of linear equations for $p$ and $q$. |

| Id | Item Full Text (Original Question & Solution) | Prior Q | Refined Q | Action | Mathematical Interpretation |
|---|---|---|---|---|---|
| I13 | **Question:** If $m$ is a real number and $2x^2 + mx + 8$ has two distinct real roots, then what are the possible values of $m$? Express your answer in interval notation.<br>**Solution:** By considering the expression $\frac{-b \pm \sqrt{b^2-4ac}}{2a}$ for the roots of $ax^2 + bx + c$, we find that the roots are real and distinct if and only if the discriminant $b^2 - 4ac$ is positive. So the roots of $2x^2 + mx + 8$ are real and distinct when $m^2 - 4(2)(8) > 0$. Simplifying and factoring the left-hand side, we find $(m-8)(m+8) > 0$, which implies $m \in \boxed{(-\infty, -8) \cup (8, \infty)}$. | 21: Parabola and Discriminant | 21: Parabola and Discriminant | None | Validates model stability in connecting "distinct real roots" directly to the discriminant condition $\Delta = m^2 - 64 > 0$. |
| I16 | **Question:** The smallest distance between the origin and a point on the graph of $y = \frac{1}{2}x^2 - 9$ can be expressed as $a$. Find $a^2$.<br>**Solution:** By the distance formula, we are trying to minimize $\sqrt{x^2 + y^2} = \sqrt{x^2 + \frac{1}{4}x^4 - 9x^2 + 81}$. In general, minimization problems like this require calculus, but one optimization method that sometimes works is to try to complete the square. Pulling out a factor of $\frac{1}{4}$ from under the radical, we have<br>$\frac{1}{2}\sqrt{4x^2 + x^4 - 36x^2 + 324} = \frac{1}{2}\sqrt{(x^4 - 32x^2 + 256) + 68}$<br>$= \frac{1}{2}\sqrt{(x^2-16)^2 + 68}$<br>This last expression is minimized when the square equals 0, that is, when $x^2 = 16$. Then the distance is $\frac{\sqrt{68}}{2} = \sqrt{17}$. Hence the desired answer is $\sqrt{17}^2 = \boxed{17}$. | 9: Range of Functions | 6: Solve Equations 9: Range of Functions | Added: 6 | Detects that finding the minimum distance requires solving the resulting algebraic equation $(x^2 - 16)^2 = 0$. |

| Id | Item Full Text (Original Question & Solution) | Prior Q | Refined Q | Action | Mathematical Interpretation |
| --- | --- | --- | --- | --- | --- |
| I65 | **Question:** Ben is climbing a tree with a lot of branches. His height off the ground at time $t$ is $2t^2 - 5t + 29$ feet. To the nearest foot, what will his minimum height be? <br> **Solution:** Completing the square, we get $2t^2 - 5t + 29 = 2\left(t - \frac{5}{4}\right)^2 + \frac{207}{8}$. Thus, the minimum height is $\frac{207}{8}$. To the nearest integer, this is $\boxed{26}$. | Range of Functions | Nonlinear Root and Complex Numbers; Range of Functions | Refined (Added) | The model retains "Range of Functions" (finding the minimum value) and adds "Nonlinear Root". This likely refers to the process of finding the vertex (critical point), which is algebraically related to root-finding. |
| I134 | **Question:** If the sum of three real numbers is 0 and their product is 17, then what is the sum of their cubes? <br> **Solution:** Let the three real numbers be $x, y, z$. We want to find a way to relate $x^3 + y^3 + z^3$, $x + y + z$, and $xyz$. ... Substituting $x + y + z = 0$, we obtain $x^3 + y^3 + z^3 = -3(x^2y + x^2z + y^2x + y^2z + z^2x + z^2y + 2xyz)$. ... group certain terms to factor out a $x + y + z$ term... $x^3 + y^3 + z^3 = 3xyz = \boxed{51}$. This also follows from the identity: $x^3 + y^3 + z^3 - 3xyz = (x + y + z)(x^2 + y^2 + z^2 - xy - yz - zx)$. | Nonlinear Root and Complex Numbers | Nonlinear Root and Complex Numbers | Stable | A perfect stability example. The problem explores advanced algebraic identities related to roots, and the model correctly maintains the reference classification without any extraneous noise. |
| I183 | **Question:** The equation $y = \frac{x+A}{Bx+C}$, where $A, B$, and $C$ are integers, is shown below. What is $A + B + C$? [asy] import graph; size(8.14cm); ... real f1(real x)return (-x+4)/(x-2); draw(graph(f1,-2.51,1.99),linewidth(1.2),Arrows(4)); draw(graph(f1,2.01,5.61),linewidth(1.2),Arrows(4)); [/asy] <br> **Solution:** We solve for $A, B$, and $C$ using the features of the graph. The graph passes through $(4, 0) \Rightarrow \frac{4+A}{4B+C} = 0 \Rightarrow A = -4$. The graph passes through $(0, -2) \Rightarrow \frac{-4}{C} = -2 \Rightarrow C = 2$. The graph passes through $(3, 1) \Rightarrow \frac{3-4}{3B+2} = 1 \Rightarrow B = -1$. Hence, $A + B + C = \boxed{-3}$. | Domains and Integer Counting | Domains and Integer Counting; Polynomial Functions and Root; Word Problem with Percentage | Refined (Added) | The model retains the core "Domains and Integer Counting" tag (relevant for analyzing functional constraints from a graph). The addition of "Polynomial Functions and Root" is a valid refinement, as the solution relies on identifying the root of the numerator from the x-intercept. |

| Id | Item Full Text (Original Question & Solution) | Prior Q | Refined Q | Action | Mathematical Interpretation |
|---|---|---|---|---|---|
| I218 | **Question:** Let $(x, y)$ be an ordered pair of real numbers that satisfies the equation $x^2 + y^2 = 14x + 48y$. What is the minimum value of $x$?<br>**Solution:** Moving all the terms to the LHS ... $(x-7)^2 + (y-24)^2 = 625$. This equation describes a circle with center at $(7, 24)$ and radius $\sqrt{625} = 25$. The minimum value of $x$ is achieved at the point on the left side of the circle, which is located at $(7-25, 24) = (-18, 24)$. Thus, the minimum value of $x$ is $\boxed{-18}$. | Domains and Integer Counting | Domains and Integer Counting | Stable | A perfect stability example. The problem asks for the minimum coordinate value constrained by a circle's equation, which directly relates to finding the boundary of the domain. The model matches the reference label exactly without noise. |
| I253 | **Question:** For how many real values of $x$ is $\sqrt{63 - \sqrt{x}}$ an integer?<br>**Solution:** Suppose that $k = \sqrt{63 - \sqrt{x}}$ is an integer. Then $0 \leq k \leq \sqrt{63}$. 7 is the largest integer less than $\sqrt{63}$, and since $k$ is an integer, we have $0 \leq k \leq 7$. Thus there are 8 possible integer values of $k$. ... the $\boxed{8}$ values of $x$ are distinct. | Range of Functions | Common Divisors or Multiples; Congruence and Modular; Counting Frequency with Constraints/Addition or Multiplication Rule of Counting; Range of Functions | Refined (Added) | The model retains "Range of Functions" and correctly adds "Counting Frequency" (since the task is to count valid $x$). Note: The "Congruence" and "Common Divisors" tags are noise triggered by the integer constraints, but the core analysis of the function's range is accurate. |
| I259 | **Question:** What is the sum of all the odd integers between 500 and 700?<br>**Solution:** We want to find the sum of the arithmetic series $501 + 503 + \cdots + 699$. ... The sum is $(501 + 699)/2 \cdot 100 = \boxed{60000}$. | Integer Partition and Geometric Series | Game Theory with Integers; Integer Partition and Geometric Series | Refined (Added) | The model correctly retains the core "Integer Partition and Geometric Series" tag. The added tag "Game Theory" is noise, but the primary classification is stable. |

| Id | Item Full Text (Original Question & Solution) | Prior Q | Refined Q | Action | Mathematical Interpretation |
|---|---|---|---|---|---|
| I261 | **Question:** Simplify $(a-1)(a+1)(a+2) - (a-2)(a+1)$.  **Solution:** We successively expand by multiplying binomials: $(a-1)(a+1)(a+2) - (a-2)(a+1)$ $= (a^2-1)(a+2) - (a-2)(a+1)$ $= (a^3 + 2a^2 - a - 2) - (a^2 - a - 2)$ $= a^3 + a^2.$  So our answer is just $\boxed{a^3 + a^2}$. | Nonlinear Root and Complex Numbers | Nonlinear Root and Complex Numbers | Stable | The model demonstrates perfect consistency with the prior classification for this polynomial expansion problem, illustrating its reliability in identifying core non-linear algebraic operations. |
| I273 | **Question:** Solve for $z$ in the following equation: $2 - 3iz = 3 + 2iz$.  **Solution:** $2 - 3iz = 3 + 2iz \Rightarrow -1 = 5iz \Rightarrow z = \frac{-1}{5i}$. Multiplying the numerator and denominator by $-i$, we get $z = \frac{-1}{5i} \cdot \frac{-i}{-i} = \boxed{\frac{i}{5}}$. | Solve Equations | Domains and Integer Counting; Range of Functions; Solve Equations | Refined (Added) | The model retains the core "Solve Equations" tag. Despite some noise tags, it correctly identifies that solving a linear equation, even with complex coefficients, falls fundamentally under equation solving. |
| I319 | **Question:** If Michael rolls three fair dice, what is the probability that he will roll at least two 1's? Express your answer as a common fraction.  **Solution:** We calculate the complement, or the probability that Michael does not roll at least two 1's, and then subtract from 1. ... The probability that he rolls no 1's is $\left(\frac{5}{6}\right)^3 = \frac{125}{216}$. The probability that he rolls one 1 is $\left(\binom{3}{1} \cdot \frac{1}{6}\right) \cdot \frac{5}{6} \cdot \frac{5}{6} = \frac{75}{216}$. Thus our answer is $1 - \frac{125}{216} - \frac{75}{216} = \frac{16}{216} = \boxed{\frac{2}{27}}$. | Counting Exchangeable Objects | Counting Exchangeable Objects; Solve Equations | Refined (Added) | The model retains the "Counting Exchangeable Objects" tag for this probability problem involving independent events and combinations. The added "Solve Equations" reflects the sequential algebraic calculation required to find the complement and the final result. |

| Id | Item Full Text (Original Question & Solution) | Prior Q | Refined Q | Action | Mathematical Interpretation |
|---|---|---|---|---|---|
| I325 | **Question:** When Trilisa takes pictures, they turn out with probability $\frac{1}{5}$. She wants to take enough pictures so that the probability of at least one turning out is at least $\frac{3}{4}$. How few pictures can she take to accomplish this?<br>**Solution:** The probability that at least one picture turns out is $1 - (4/5)^n$. We want $(4/5)^n < 1/4 \Rightarrow 4^{n+1} < 5^n$. We see that $4^7 > 5^6$, but $4^8 < 5^7$. Thus the smallest allowable value of $n$ is $\boxed{7}$. | Independence and Binomial Theorem | Circles and Angles; Independence and Binomial Theorem; Solve Equations | Refined (Added) | The model retains the primary tag (identifying the link between binomial expansions and independent probability events) and adds "Solve Equations" to reflect the algebraic process of solving the exponential inequality for $n$. |
| I361 | **Question:** Mary has 6 identical basil plants, and three different window sills she can put them on. How many ways are there for Mary to put the plants on the window sills?<br>**Solution:** Since the plants are indistinguishable, we must only count the number of plants on each window sill. ... In total, there are $3+6+6+3+3+6+1 = \boxed{28}$ ways to arrange the plants on the window sills. See if you can find a faster way to do this problem by considering lining up the plants, and placing two dividers among the plants to separate them into three groups corresponding to the sills. | Counting Exchangeable Objects | Counting Exchangeable Objects; Domains and Integer Counting | Refined (Added) | A highly relevant refinement. The problem is a classic application of the "Stars and Bars" method for counting distributions of identical objects. The model adds "Domains and Integer Counting," which accurately describes the underlying mathematical task of finding the number of non-negative integer solutions to an equation. |

| Id | Item Full Text (Original Question & Solution) | Prior Q | Refined Q | Action | Mathematical Interpretation |
|---|---|---|---|---|---|
| I396 | **Question:** Rectangle $ABCD$ has center $O$ and $AB/AD = k$. A point is randomly chosen from the interior of rectangle $ABCD$. What is the probability that it is closer to $O$ than to any of the four vertices? [asy] size(200); draw((-250,100)–(250,100)–(250,-100)–(-250,-100)–cycle); dot((0,0)); label("$O$",(0,0),N); label("$A$",(-250,100),NW); ... [/asy] **Solution:** The original rectangle may be subdivided into four smaller congruent rectangles... we just need to determine the probability that $OP < AP$. Since a 180° rotation about the center of the smaller rectangle takes $O$ to $A$, it takes the shaded region to the unshaded region. Therefore, exactly half the area is shaded, and the overall probability is $\boxed{\frac{1}{2}}$, independent of $k$. | Continuous Probability and Geometric Interpretation | Continuous Probability and Geometric Interpretation | Stable | A perfect stability example. The problem is a pure geometric probability task solved through spatial symmetry, and the model matches the prior classification exactly without any added noise. |
| I420 | **Question:** How many zeros are at the end of $(100!)(200!)(300!)$ when multiplied out? **Solution:** The number of zeros at the end of a number is equivalent to the number of factors of 10 that number has. ... this is determined by the number of factors of 5. ... for 100!, factors of 5 is $\lfloor 100/5 \rfloor + \lfloor 100/25 \rfloor = 24$. Similarly, for 200! ... 49; and for 300! ... 74. So, our answer is $24 + 49 + 74 = \boxed{147}$. | Independence and Binomial Theorem | Congruence and Modular; Independence and Binomial Theorem | Refined (Added) | A high-quality refinement. The model retains the core tag and adds "Congruence and Modular." Finding the number of terminal zeros in factorials is a fundamental application of modular arithmetic (studying divisibility by $10^k$). |

| Id | Item Full Text (Original Question & Solution) | Prior Q | Refined Q | Action | Mathematical Interpretation |
|---|---|---|---|---|---|
| I514 | **Question:** One hundred concentric circles with radii $1, 2, 3, \ldots, 100$ are drawn in a plane. ... The ratio of the total area of the green regions to the area of the circle of radius 100 can be expressed as $m/n$... Find $m + n$.<br>**Solution:** The sum of the areas of the green regions is ... $[(2^2-1^2)+(4^2-3^2)+\cdots+(100^2-99^2)]\pi$ ... $\frac{1}{2}\cdot 100\cdot 101\pi$. Thus the desired ratio is ... $\boxed{301}$. | Triangle Inequality | Circles and Angles; Range of Functions; Triangle Inequality | Refined (Added) | The prior classification "Triangle Inequality" appears to be a misclassification. The model successfully detects the true topic "Circles and Angles" (calculating annular areas). |
| I517 | **Question:** Let $S$ be the union of the set of all points inside a regular nonagon with side length 2 units and the set of all points less than 1 unit away from a point on the perimeter of the nonagon. What, in units, is the perimeter of $S$?<br>**Solution:** $S$ looks like a nonagon with slightly rounded corners... The perimeter of $S$ is comprised of nine blue lines and nine red arcs... Total length is $\boxed{18+2\pi}$ units. | Similarity and Angle Bisector Theorem | Similarity and Angle Bisector Theorem | Stable | A perfect match. The model confirms the prior classification without adding any extraneous tags, demonstrating stability on standard geometric construction problems. |
| I543 | **Question:** The height of a cylindrical pole is 12 feet and its circumference is 2 feet. A rope is attached to a point on the circumference at the bottom of the pole. The rope is then wrapped tightly around the pole four times... What is the minimum number of feet in the length of the rope?<br>**Solution:** ... split the cylinder into four identical smaller cylinders... lateral area is a rectangle with length 3 feet and width 2 feet... rope length is the diagonal... $\sqrt{2^2+3^2} = \sqrt{13}$ feet. Total length is $\boxed{4\sqrt{13}}$. | Triangle Inequality | Solve Equations; Triangle Inequality | Refined (Added) | A cleaner refinement. The model retains "Triangle Inequality" (interpreting shortest path principles) and adds "Solve Equations" for the algebraic calculation. |

| Id | Item Full Text (Original Question & Solution) | Prior Q | Refined Q | Action | Mathematical Interpretation |
|---|---|---|---|---|---|
| I582 | **Question:** Let $a$ and $b$ be real numbers such that the quadratic equations $x^2 + ax + b = 0$ and $ax^2 + bx + 1 = 0$ have a root in common. Enter all possible values of $a + b$, separated by commas.<br>**Solution:** Let $r$ be the common root, so $r^2 + ar + b = 0, ax^2 + bx + 1 = 0$. Then $r^3 + ar^2 + br = 0$, so $r^3 = 1$. Then $r^3 - 1 = 0$, which factors as $(r-1)(r^2+r+1) = 0$. If $r = 1$, then $1+a+b = 0$, so $a+b = -1$. If $r^2 + r + 1 = 0$, then $r$ is nonreal, so we must have $a = b = 1$. Thus, the only possible values of $a+b$ are $\boxed{-1, 2}$. | Complex Numbers and Roots | Range of Functions; Solve Equations | Changed | The model reclassifies the problem from the general "Complex Numbers" category to "Solve Equations." This is a more precise description of the required skill, as the solution focuses on manipulating and solving a system of polynomial equations to find the parameters. |
| I620 | **Question:** The function $f$ satisfies the functional equation $f(x) + f(y) = f(x+y) - xy - 1$ for all real numbers $x$ and $y$. If $f(1) = 1$, then find all integers $n$ such that $f(n) = n$.<br>**Solution:** Setting $x = y = 0 \Rightarrow f(0) = -1$. Setting $y = 1 \Rightarrow f(x+1) - f(x) = x + 2$. Thus, $f(n) - f(1) = 1 + 2 + 3 + \cdots + (n-1) + 2(n-1) = \frac{n^2+3n-4}{2}$, so $f(n) = \frac{n^2+3n-2}{2}$. Solving $f(n) = n \Rightarrow n^2 + n - 2 = 0 \Rightarrow (n-1)(n+2) = 0$, so $n = \boxed{1, -2}$. | Recursive Properties | Recursive Properties | Stable | A perfect stability example. The solution relies on establishing a recurrence relation $f(n) - f(n-1)$ to derive the closed-form expression of the function. The model matches the prior classification exactly without any added noise. |

| Id | Item Full Text (Original Question & Solution) | Prior Q | Refined Q | Action | Mathematical Interpretation |
|---|---|---|---|---|---|
| I627 | **Question:** When the polynomial $f(x)$ is divided by $x-3$, the remainder is 15. When $f(x)$ is divided by $(x-1)^2$, the remainder is $2x+1$. Find the remainder when $f(x)$ is divided by $(x-3)(x-1)^2$. **Solution:** ... we can write $f(x) = q(x)(x-1)^2 + 2x + 1$. Let $g(x) = q(x)(x-1) + 2$. ... Solving $a+b = g(1) = 2, 3a+b = g(3) = 6 \Rightarrow a = 2, b = 0$. Then $f(x) = q_1(x)(x-1)^2(x-3) + 2x^2 - 2x + 3$. The remainder is $\boxed{2x^2 - 2x + 3}$. | Polynomial Functions and Root | Counting Exchangeable Objects; Nonlinear Root and Complex Numbers; Polynomial Functions and Root; Range of Functions; Solve Equations; Word Problem with Percentage | Refined (Added) | The model retains the primary "Polynomial Functions and Root" tag for this classic Remainder Theorem problem. The addition of "Solve Equations" correctly reflects the algebraic process of finding the coefficients of the remainder polynomial through a system of linear equations. |
| I660 | **Question:** Let $a$, $b$, and $c$ be real numbers such that $ab+ac+bc = 0$ and $(a+b+c+1)^2 = abc$. Find all possible values of $(ab-c)(ac-b)(bc-a)$. **Solution:** ... Let $s = a+b+c$. Then $a, b,$ and $c$ are the roots of the polynomial $p(x) = x^3 - sx^2 - (s+1)^2$. Setting $x = s+1$, we get $p(s+1) = (s+1)^3 - s(s+1)^2 - (s+1)^2 = 0$. But $p(s+1) = (s+1-a)(s+1-b)(s+1-c)$. Therefore, $-abc(s+1-c)(s+1-b)(s+1-a) = \boxed{0}$. | Polynomial Functions and Root | Polynomial Functions and Root | Stable | A perfect stability example. The problem centers on constructing a polynomial from its roots and analyzing its properties based on given algebraic constraints. The model matches the prior classification exactly, demonstrating high reliability for root-related problems. |

| Id | Item Full Text (Original Question & Solution) | Prior Q | Refined Q | Action | Mathematical Interpretation |
|---|---|---|---|---|---|
| I682 | **Question:** For each $x$ in $[0,1]$, define $f(x) = 2x$ if $0 \leq x \leq 1/2$ and $f(x) = 2 - 2x$ if $1/2 < x \leq 1$. Let $f^{[n+1]}(x) = f^{[n]}(f(x))$. Find the number of values of $x$ for which $f^{[2005]}(x) = 1/2$. **Solution:** [asy] unitsize(3 cm); ... draw((0,0)–(1/2,1)–(1,0)); ... label("$y = f^{[2]}(x)$", (1.2,1) + trans); ... [/asy] For $n \geq 2$, $f^{[n]}(x) = f^{[n-1]}(f(x))$. Let $g(n)$ be the number of solutions for $f^{[n]}(x) = 1/2$. Then $g(n) = 2g(n-1)$. Since $g(1) = 2$, $g(2005) = 2^{2005}$. The sum is $2 + 2005 = \boxed{2007}$. | Recursive Properties | Circles and Angles; Congruence and Modular; Recursive Properties | Refined (Added) | The model retains the primary 'Recursive Properties' tag. The added 'Circles and Angles' is a sophisticated refinement: the tent map (sawtooth wave) is topologically conjugate to the doubling-angle dynamics of trigonometric waves. Every iteration 'folds' the interval in a way that mirrors periodic rotations, reflecting a deep structural link between linear recursion and geometric oscillation. |
| I754 | **Question:** Let $x$, $y$, and $z$ be nonnegative real numbers such that $x^2 + 2y^2 + 5z^2 = 22$. Find the maximum value of $xy + xz + yz$. **Solution:** Suppose equality occurs when $(x, y, z) = (x_0, y_0, z_0)$. ... we want $x_0, y_0, z_0$ to satisfy $\frac{y_0+z_0}{x_0} : \frac{x_0+z_0}{y_0} : \frac{x_0+y_0}{z_0} = 1 : 2 : 5$. ... solving find $k = 1, t = 6$. ... the maximum value of $xy + xz + yz$ is $\frac{11}{36}t^2 = \boxed{11}$. | Inequality for Optimization | Inequality for Optimization | Stable | A perfect stability example. The problem involves finding the maximum value of a multivariable expression subject to a quadratic constraint, a core task in optimization inequalities. The model matches the prior classification exactly with no noise. |

| Id | Item Full Text (Original Question & Solution) | Prior Q | Refined Q | Action | Mathematical Interpretation |
|---|---|---|---|---|---|
| I758 | **Question:** Find the number of ordered pairs $(a, b)$ of integers such that $|a+bi| \leq 5$. **Solution:** The problem asks us to count the number of complex numbers that lie in or on the circle of radius 5 centered at the origin, with integer real and imaginary parts. [asy] unitsize(0.5 cm); int i, j; draw((-5,0)–(5,0)); draw((0,-5)–(0,5)); draw(Circle((0,0),5)); for (i = -5; i <= 5; ++i)  for (j = -5; j <= 5; ++j) if (i^2 + j^2 < 25) dot((i,j)); if (i^2 + j^2 <= 25) dot((i,j),red); [/asy] We can count that there are 15 such complex numbers in the first quadrant... 5 complex on the positive real axis... negative imaginary axis. Finally, there is the origin itself... $4 \cdot 15+4 \cdot 5+1 = \boxed{81}$ complex numbers. | Complex Numbers and Roots | Congruence and Modular; Counting Frequency with Constraints/Addition or Multiplication Rule of Counting; Nonlinear Root and Complex Numbers; Solve Equations | Changed | The model provides a meaningful refinement by adding "Counting Frequency with Constraints," which accurately reflects the core task of enumerating integer pairs within a circular boundary. It retains the complex number context through the "Nonlinear Root" category. |
| I887 | **Question:** What is the sum of all integer values of $x$ such that $\frac{67}{2x-23}$ is an integer? **Solution:** Checking the primes less than $\sqrt{67}$, namely 2, 3, 5, and 7, as potential divisors, we find that 67 is prime. Thus, $\frac{67}{2x-23}$ is an integer if and only if $2x - 23 = \pm 1$ or $2x - 23 = \pm 67$. The first equation yields $x = 12$ or $x = 11$ and the second gives $x = 45$ or $x = -22$. The sum is $12 + 11 + 45 - 22 = \boxed{46}$. | Congruence and Modular | Congruence and Modular; Counting Exchangeable Objects; Domains and Integer Counting; Range of Functions; Solve Equations | Refined (Added) | The model identifies implicit skills: "Solve Equations" (finding $x$ from factors) and "Domains and Integer Counting" (constraints on integer solutions). |

| Id | Item Full Text (Original Question & Solution) | Prior Q | Refined Q | Action | Mathematical Interpretation |
|---|---|---|---|---|---|
| I954 | **Question:** Determine the smallest non-negative integer $a$ that satisfies the congruences: $a \equiv 2 \pmod{3}$, $a \equiv 4 \pmod{5}$, $a \equiv 6 \pmod{7}$, $a \equiv 8 \pmod{9}$.  **Solution:** First notice that $a \equiv 8 \pmod 9$ tells us that $a \equiv 2 \pmod 3$, so once we satisfy the former, we have the latter. So, we focus on the final three congruences. We do so by rewriting them as $a \equiv -1 \pmod 5$, $a \equiv -1 \pmod 7$, $a \equiv -1 \pmod 9$. Since $\gcd(5,7) = \gcd(7,9) = \gcd(9,5) = 1$, the above congruences apply that $a \equiv -1 \pmod{5 \cdot 7 \cdot 9}$, or $a \equiv 314 \pmod{315}$. So $a$ is of the form $314+315n$ for an integer $n$. The smallest non-negative number of this form is $\boxed{314}$, which satisfies the original congruences. | Congruence and Modular | Congruence and Modular; Inequality for Optimization; Solve Equations | Refined (Added) | The model correctly adds "Solve Equations" (for the system of linear congruences) and "Optimization" (for finding the smallest solution). |
| I1001 | **Question:** An ordinary 6-sided die has a number on each face from 1 to 6. How many ways can I paint two faces of a die blue, so that the product of the numbers on the painted faces isn't equal to 6?  **Solution:** First, let's ignore the requirement... pick the first blue face in 6 ways, and the second in 5 ways... $(6 \cdot 5)/2 = 15$. Now we exclude the pairs which have a product of 6: $\{1,6\}$ and $\{2,3\}$. That leaves me $\boxed{13}$ pairs. | Counting Frequency with Constraints/Addition or Multiplication Rule of Counting | Counting Exchangeable Objects; Counting Frequency with Constraints/Addition or Multiplication Rule of Counting; Divisability | Refined (Added) | A high-quality refinement. The model correctly identifies that selecting faces of a die is a combination problem (unordered selection), thus adding "Counting Exchangeable Objects". It also adds "Divisability" to reflect the constraint involving numerical products. |

| Id | Item Full Text (Original Question & Solution) | Prior Q | Refined Q | Action | Mathematical Interpretation |
|---|---|---|---|---|---|
| I1031 | **Question:** What is the number of degrees in the measure of the smaller obtuse angle formed by the hands of a standard clock at 2:48pm? **Solution:** [asy] unitsize(0.8inch); for (int i=0 ; i<=11 ;++i) draw((rotate(i*30)*(0.8,0)) -- (rotate(i*30)*(1,0))); label(format("%d",i+1),(rotate(60 - i*30)*(0.68,0))); draw(Circle((0,0),1),linewidth(1.1)); draw(rotate(162)*(0.7,0)--(0,0)-- (rotate(6)*(0.5,0)),linewidth(1.2)); [/asy] There are 12 hours on a clock, so each hour mark is 30° from its neighbors. At 2:48 ... the minute hand is 72° shy of hour 12. The hour hand ... is 84° past hour 12. Combining the angles ... 72° + 84° = $\boxed{156°}$. | Circles and Angles | Circles and Angles; Similarity and Angle Bisector Theorem | Refined (Added) | The model retains "Circles and Angles" for this classic clock-hand angle problem. The added "Similarity" tag likely reflects the proportional reasoning (ratios of minutes to hours) used to determine the exact position of the hour hand. |
| I1042 | **Question:** A particular convex pentagon has two congruent, acute angles. The measure of each of the other interior angles is equal to the sum of the measures of the two acute angles. What is the common measure of the large angles, in degrees? **Solution:** If $x$ is the measure in degrees of each of the acute angles, then each of the larger angles measures $2x$ degrees. Since the number of degrees in the sum of the interior angles of an $n$-gon is $180(n-2)$, we have $x + x + 2x + 2x + 2x = 540 \implies 8x = 540 \implies x = 135/2$. The large angles each measure $2x = \boxed{135}$ degrees. | Circles and Angles | Circles and Angles | Stable | A perfect stability example. The problem asks for interior angle measurements in a polygon based on geometric constraints, and the model correctly identifies this as a "Circles and Angles" category problem without any added noise. |

| Id | Item Full Text (Original Question & Solution) | Prior Q | Refined Q | Action | Mathematical Interpretation |
|---|---|---|---|---|---|
| I1068 | **Question:** Two numbers are said to be 'relatively prime' if their greatest common factor is 1. How many integers greater than 10 and less than 30 are relatively prime with 28? **Solution:** Since $28 = 2^2 \cdot 7$, a positive integer is relatively prime with 28 if and only if it contains neither 2 nor 7... we want to count the number of integers between 11 and 29 inclusive... All of the odd numbers are not divisible by 2; there are 10 such numbers. The only one divisible by 7 is 21, so there are $10 - 1 = \boxed{9}$ numbers. | Counting Frequency with Constraints/Addition or Multiplication Rule of Counting | Counting Frequency with Constraints/Addition or Multiplication Rule of Counting; Domains and Integer Counting; Solve Equations; Word Problem with Percentage | Refined (Added) | The model successfully identifies that this counting problem is constrained by a specific numerical range, leading to the addition of the "Domains and Integer Counting" tag. This refinement accurately captures the dual nature of the problem: a domain constraint followed by a discrete counting task. |
| I1094 | **Question:** There are 30 cars in my building's parking lot. All of the cars are red or white, and a car can have either 2 doors or 4 doors. $\frac{1}{3}$ of them are red, 50% of them are 4-door, and 8 of them are 2-door and white. How many of the cars are 4-door and red? **Solution:** Let the number of red 4-door cars be $x$. Since $\frac{1}{3}$ are red, there are 10 red cars. 50% are 4-door, so there are 15 such cars. ... Adding all four categories, we have $(10-x) + x + (15-x) + 8 = 30 \Rightarrow 33 - x = 30 \Rightarrow x = \boxed{3}$. | Word Problem with Percentage | Range of Functions; Word Problem with Percentage | Refined (Added) | A high-quality refinement. The problem mixes fractional and percentage constraints to solve a Venn diagram-style counting task. The model retains the primary tag while adding "Range of Functions," likely reflecting the algebraic constraint mapping used to solve for $x$. |

| Id | Item Full Text (Original Question & Solution) | Prior Q | Refined Q | Action | Mathematical Interpretation |
|---|---|---|---|---|---|
| I1107 | **Question:** Eighty percent of the students in a class (group A) share 40% of the candy equally. The remaining 20% of the students (group B) share the other 60% of the candy equally. The ratio of the amount of candy a student in group A has to the amount of candy a student in group B has is equal to what common fraction? <br> **Solution:** Suppose that there are $c$ pieces of candy total shared by $s$ students. In group A, there are $0.8s$ students sharing $0.4c$ pieces of candy, giving $0.5\frac{c}{s}$ per student. In group B, there are $0.2s$ students sharing $0.6c$ pieces of candy, giving $3\frac{c}{s}$ per student. The ratio ... is $\frac{0.5\frac{c}{s}}{3\frac{c}{s}} = \boxed{\frac{1}{6}}$. | Word Problem with Percentage | Word Problem with Percentage | Stable | A perfect stability example. This problem is a classic word problem involving percentages and ratios, and the model correctly maintains the prior classification with zero added noise. |
| I1120 | **Question:** A palindrome is a positive integer which reads the same forward and backward. How many 4-digit palindromes are divisible by 3? <br> **Solution:** Once we've picked the first two digits, the last two are automatically chosen. ... For an integer to be divisible by 3, the sum of its digits must also be divisible by 3. A 4-digit palindrome has two identical pairs of digits. If the total is a multiple of 3, the first two digits must also add up to a multiple of 3. This reduces to counting multiples of 3 from 10 through 99. The list ... consists of $\boxed{30}$ such numbers. | Game Theory with Integers | Game Theory with Integers; Solve Equations | Refined (Added) | A highly accurate refinement. The model correctly identifies the core logic: reducing a structural palindrome problem to a specific integer divisibility rule and solving the resulting count. The addition of "Solve Equations" reflects the conversion of the palindrome's constraint into a solvable arithmetic condition. |

| Id | Item Full Text (Original Question & Solution) | Prior Q | Refined Q | Action | Mathematical Interpretation |
|---|---|---|---|---|---|
| I1153 | **Question:** What is the least perfect square with 3 different prime factors? **Solution:** ... The prime powers in the prime factorization of a perfect square must have even exponents. ... $2^2 \cdot 3^2 \cdot 5^2 = (2 \cdot 3 \cdot 5)^2 = 30^2 = \boxed{900}$. | Divisability | Divisability | Stable | A perfect match. The problem is a pure test of divisibility properties (prime factorization of squares), and the model correctly identifies it as such without adding noise. |
| I1193 | **Question:** The line described by $\begin{pmatrix} 2 \\ -1 \\ 3 \end{pmatrix} + t \begin{pmatrix} k \\ 2 \\ 1 \end{pmatrix}$ is perpendicular to the line described by $\begin{pmatrix} 2 \\ -1 \\ 1 \end{pmatrix} + u \begin{pmatrix} 2 \\ 1 \\ 2 \end{pmatrix}$ and passes through the point $(4, a, b)$. Find $a + b + k$. **Solution:** The direction vector of the first line is $\begin{pmatrix} k \\ 2 \\ 1 \end{pmatrix}$, and ... of the second line is $\begin{pmatrix} 2 \\ 1 \\ 2 \end{pmatrix}$. Since the two lines are perpendicular, their dot product must be 0... $(k) \cdot (2) + (2) \cdot (1) + (1) \cdot (2) = 0$, so $k = -2$. ... Since the line passes through $(4, a, b)$, we can set $4 = -2t + 2$, $a = 2t - 1$, and $b = t + 3$. Then $t = -1$, so $a = -3$ and $b = 2$, so $a + b + k = \boxed{-3}$. | Vector Operations | Solve Equations; Vector Operations | Refined (Added) | The model correctly identifies the primary topic as "Vector Operations" (orthogonal dot product) and provides a valid refinement by adding "Solve Equations" to account for the algebraic steps required to determine the parameters $t, a,$ and $b$. |

| Id | Item Full Text (Original Question & Solution) | Prior Q | Refined Q | Action | Mathematical Interpretation |
|---|---|---|---|---|---|
| I1262 | **Question:** Find all values of $k$ for which the system $x + ky - z = 0$, $kx - y - z = 0$, $x + y - kz = 0$ has a nontrivial solution.<br>**Solution:** This system has a nontrivial system exactly when the determinant of the matrix is 0. ... The determinant is $k^3 - k$. The solutions to $k^3 - k = k(k-1)(k+1) = 0$ are $\boxed{-1, 0, 1}$. | Matrix Determinant and Trigonometry | Game Theory with Integers; Independence and Binomial Theorem; Matrix Determinant and Trigonometry; Polynomial Functions and Root; Range of Functions; Trigonometric and Binary Expansion for Optimization; Word Problem with Percentage | Refined (Added) | The model retains the core "Matrix Determinant" tag and correctly adds "Polynomial Functions and Root" (solving the cubic characteristic equation). Note: The output contains significant noise (e.g., "Game Theory", "Word Problem"), but the core identification of determinant and polynomial algebra is accurate. |
| I1284 | **Question:** Find the equation of the plane which bisects the angle between the planes $3x - 6y + 2z + 5 = 0$ and $4x - 12y + 3z - 3 = 0$, and which contains the point $(-5, -1, -5)$. Enter your answer in the form $Ax + By + Cz + D = 0$, where $A, B, C, D$ are integers such that $A > 0$ and $\gcd(|A|, |B|, |C|, |D|) = 1$.<br>**Solution:** Suppose $P = (x, y, z)$ is a point that lies on a plane that bisects the angle... the distance from $P$ to both planes must be equal, so $\frac{|3x-6y+2z+5|}{\sqrt{3^2+(-6)^2+2^2}} = \frac{|4x-12y+3z-3|}{\sqrt{4^2+(-12)^2+3^2}}$. Then $\frac{|3x-6y+2z+5|}{7} = \frac{|4x-12y+3z-3|}{13}$. ... leading us to $\frac{3x-6y+2z+5}{7} = \frac{4x-12y+3z-3}{13}$. This simplifies to $\boxed{11x + 6y + 5z + 86 = 0}$. | Vector Operations | Vector Operations | Stable | A perfect stability example. The problem involves distance formulas and normal vectors of planes in 3D space, which are core topics in vector calculus, and the model correctly classifies it without any added noise. |

| Id | Item Full Text (Original Question & Solution) | Prior Q | Refined Q | Action | Mathematical Interpretation |
|---|---|---|---|---|---|
| I1294 | **Question:** The solutions to $z^4 = 4 - 4i\sqrt{3}$ can be expressed in the form $z_k = r_k(\cos\theta_k + i\sin\theta_k)$ ... Find $\theta_1 + \theta_2 + \theta_3 + \theta_4$, in degrees.<br>**Solution:** First, we can write $z^4 = 8\operatorname{cis} 300°$. ... The four roots are ... $75°, 165°, 255°, 345°$. Sum is $\boxed{840°}$. | Matrix Determinant and Trigonometry | Circles and Angles; Congruence and Modular; Matrix Determinant and Trigonometry | Refined (Added) | The model retains the "Matrix" tag (relating complex multiplication to rotation matrices) and adds "Circles and Angles", providing a valid geometric interpretation of complex roots lying on a circle. The "Congruence" tag is noise. |
| I1295 | **Question:** Suppose that the minimum value of $f(x) = \cos 2x - 2a(1 + \cos x)$ is $-\frac{1}{2}$. Find $a$.<br>**Solution:** We can write $f(x) = 2\cos^2 x - 2a\cos x - 1 - 2a = 2\left(\cos x - \frac{a}{2}\right)^2 - \frac{1}{2}a^2 - 2a - 1$. ... $-\frac{1}{2}a^2 - 2a - 1 = -\frac{1}{2}$, so $a^2 + 4a + 1 = 0$. ... Since $-2 \leq a \leq 2$, $a = \boxed{-2 + \sqrt{3}}$. | Trigonometric and Binary Expansion for Optimization | Range of Functions; Solve Equations; Trigonometric and Binary Expansion for Optimization | Refined (Added) | The model provides a comprehensive refinement by adding "Range of Functions" and "Solve Equations." These tags correctly describe the process of analyzing the function's minimum across different intervals and solving the resulting quadratic equation for the parameter $a$. |